\documentclass[12pt]{emulateapj}

\usepackage{longtable,natbib}

\begin{document}

\title{Insights into the Cepheid distance scale}

\shorttitle{Insights into the Cepheid distance scale.} 
\shortauthors{Bono et al.}

\author{
G. Bono\altaffilmark{1,2},
F. Caputo\altaffilmark{2},
M. Marconi\altaffilmark{3} and 
I. Musella\altaffilmark{3}  
} 

\altaffiltext{1}{Dipartimento di Fisica, Universit\`a di Roma Tor Vergata, 
via della Ricerca Scientifica 1, 00133 Rome, Italy; bono@roma2.infn.it}

\altaffiltext{2}{INAF -- Osservatorio Astronomico di Roma, Via Frascati 33, 
00040 Monte Porzio Catone, Italy; caputo@oa-roma.inaf.it}

\altaffiltext{3}{INAF -- Osservatorio Astronomico di Capodimonte, 
Via Moiariello 16, 80131 Napoli, Italy; marcella.marconi@oacn.inaf.it, ilaria.musella@oacn.inaf.it}

\date{\centering drafted \today\ / Received / Accepted }

\begin{abstract}
We present a detailed investigation of the Cepheid distance scale by using 
both theory and observations. Through the use of pulsation models for fundamental 
mode Cepheids, we found that the slope of the Period-Luminosity ($P$-$L$) relation 
covering the entire period range (0.40$\le$log$P\le$2.0) becomes steeper 
when moving from optical to near-infrared (NIR) bands, and that the 
metallicity dependence of the slope decreases from the $B$- to the $K$-band.
The sign of the metallicity dependence for the slopes of the $P$-$L_V$ and 
$P$-$L_I$ relation is at odds with some recent empirical estimates. 

We determined new homogeneous estimates of $V$- and $I$-band slopes
for 87 independent Cepheid data sets belonging to 48 external galaxies 
with nebular oxygen abundance 7.5$\le$12+$\log$(O/H)$\le$8.9.
By using Cepheid samples including more than 20 Cepheids,
the $\chi^2$ test indicates that the hypothesis of a steepening of the 
$P$-$L_{V,I}$ relations with increased metal content can be discarded 
at the 99\% level. On the contrary, the observed slopes agree with the 
metallicity trend predicted by pulsation models, i.e. the slope is 
roughly constant for galaxies with 12+$\log$(O/H)$<$8.17 and becomes 
shallower in the metal-rich regime, with a confidence level of 62\% 
and 92\%, respectively. The $\chi^2$ test concerning the hypothesis 
that the slope does not depend on metallicity gives confidence levels 
either similar ($PL_V$, 62\%) or smaller ($PL_I$, 67\%).

We investigated the dependence of the Period-Wesenheit ($P$-$W$) 
relations on the metal content and we found that the slopes of optical 
and NIR $P$-$W$ relations in external galaxies are similar to the  
slopes of Large Magellanic Cloud (LMC) Cepheids. They also 
agree with the theoretical predictions suggesting that the slopes 
of the $P$-$W$ relations are independent of the metal content. 
On this ground, the $P$-$W$ relations provide a robust method to 
determine distance moduli relative to the LMC, but theory and 
observations indicate that the metallicity dependence of the 
zero-point in the different passbands has to be taken into account. 

To constrain this effect, we compared the independent set of galaxy 
distances provided by Rizzi et al.\ (2007) using the Tip of the 
Red Giant Branch (TRGB) with our homogeneous set of extragalactic 
Cepheid distances based on the $P$-$W$ relations. We found that 
the metallicity correction on distances based on the $P$-$WBV$ relation 
is $\gamma_{B,V}$=$-0.52$ mag dex$^{-1}$, whereas it is vanishing for 
the distances based on the $P$-${WVI}$ and on the $P$-${WJK}$ relations. 
%Point C 
These findings fully support Cepheid theoretical predictions.
\end{abstract}


\maketitle
%
%________________________________________________________________

\pagebreak

%%%%%%%%%%%%%%%%%%%%%%%%%%%%%%%%%%%%%%%%%%%%%%%%%%%%%%%%%%%%%%%%%%%%%%%%%%%
\section{Introduction}

The cosmic distance scale and the estimate of the Hubble constant, $H_0$, 
are tightly connected with the Period-Luminosity ($P$-$L$) relation of 
Classical Cepheids and the distances to external galaxies are traditionally 
determined by using a universal $P$-$L$ linear relation based on the 
Large Magellanic Cloud (LMC) variables.

However, theoretical predictions based on nonlinear, convective Cepheid 
models computed by our group (see Caputo 2008 for a comprehensive list 
of references)
indicate that the instability strip boundaries in the log$L$-log$T_e$ 
plane are almost linear, but when transformed into the different 
Period-Magnitude planes they are better described by quadratic 
$P$-$L$ relations. 
In particular, we found that at fixed metal content: (1) the
predicted optical $P$-$L$ relations can be properly fit with 
quadratic relations, (2) a discontinuity around
log$P\sim$ 1.2 should be adopted to constrain the theoretical
results into linear approximations, and (3) the predicted $P$-$L$
relations become more and more linear and tight when moving from
optical to near-infrared bands. Moreover, we suggested that the
metal-poor Cepheids follow $P$-$L$ relations which are steeper and
brighter than the metal-rich ones, with the amount of this
metallicity effect again decreasing from the $B$ to the $K$ band.
Furthermore, we drew attention to the evidence that the metallicity
effect on the predicted Period-Wesenheit ($P$-$W$) relations, which
present several advantages when compared with the $P$-$L$ relations,
significantly depends on the adopted Wesenheit function.

With a few exceptions, these theoretical results have been considered 
with a certain skepticism. Only during the last few years,   
several observational investigations disclosed the nonlinearity 
of the $P$-$L$ relation (Tammann, Sandage, \& Reindl 2003; 
Ngeow et al.\ 2005; Ngeow \& Kanbur 2006), as well as the evidence that 
the Cepheid $P$-$L$ relation cannot be universal and that both the slope 
and the zero-point might change from galaxy to galaxy. Quoting
Sandage, Tammann \& Reindl (2009), "{\it the existence of a
universal $P$-$L$ relation is an only historically justified
illusion".}

According to these new empirical evidence, which might imply
severe limits in the precision of Cepheid distances, we examine 
the available observations by using the theoretical framework provided 
by the pulsation models and we address three main issues concerning 
the Cepheid distance scale: the intrinsic features of the $P$-$L$
and the $P$-$W$ relations, the dependence of the $P$-$L$ and of the 
$P$-$W$ slope on the Cepheid metal content and the impact of the 
metallicity effect on the Cepheid distances.

%%%%%%%%%%%%%%%%%%%%%%%%%%%%%%%%%%%%%%%%%%%%%%%%%%%%%%%%%%%%%%%%%%%%%%
\section{Intrinsic features of the P-L relation}

\subsection{Predictions based on pulsation models}

The theoretical framework we developed during the last ten years was 
already described in a series of papers (see e.g. Caputo 2008; Marconi 2009) 
and its main lines can be summarized as follows.

{\em i)} A nonlinear, nonlocal, and time-dependent convective code is 
used to calculate several model sequences (see Table 1) with constant chemical
composition, mass and luminosity, by varying the effective temperature $T_e$ 
with steps of 100 K in order to properly cover the pulsation region.  
For metal abundances $Z\ge$0.02, the adopted helium content $Y$ accounts 
for different values of the helium-to-metals enrichment ratio 
$\Delta Y/\Delta Z$. For each chemical composition and mass value, the 
luminosity is fixed using a Mass-Luminosity ($ML$) relation based on 
evolutionary models that either neglect (``canonical'') or account for 
convective core overshooting during central hydrogen burning phases  
(``overshooting'').  the canonical $ML$ relation adopted to construct 
our pulsation models is that given by Bono et al.\ (2000): \par 

$$\log L_{can}=0.90+3.35\log M+1.36\log Y-0.34\log Z\eqno(1)$$\par 

\noindent 
which has a standard deviation $\sigma$=0.02 and accounts for both 
the blueward and the redward crossing of the instability strip. 
Note  that, together with minor discrepancies among the
canonical $ML$ relations provided by different authors (see also
Girardi et al.\ 2000; Castellani et al.\ 2003; Bertelli et al.\ 2009;   
Valle et al.\ 2009, and references therein),
a variation in luminosity with respect to the canonical level can
be due either to convective core overshooting or to mass-loss before 
or during the He-burning phases. In the former case, stellar structures 
at fixed mass and chemical composition are over-luminous by 
log$L/L_{can}$=0.25 (see Chiosi, Wood, \& Capitanio 1993), while in 
the latter the stellar structures at fixed luminosity and chemical 
composition are less massive (see Castellani \& Degl'Innocenti 1995; 
Bono et al.\ 2000; Castellani et al.\ 2003). The net consequence 
of the quoted physical mechanisms is a positive log$L/L_{can}$ ratio.
Finally, all the pulsation models are calculated assuming a mixing-length 
parameter\footnote{ The mixing-length parameter is a measure of the convection
efficiency. It is used to close the system of equations describing
the dynamical and convective stellar structure.} $l/H_p$=1.5, but
several additional models were computed by adopting $l/H_p$=1.7 
and 1.8.

{\em ii)} The adopted approach provides not only the pulsation 
equation $P$=$f(Z,Y,M,L,T_e)$ and the blue (hot) boundaries of the 
instability strip, but also robust predictions concerning the 
red (cool) boundaries together with the pulsation amplitudes 
(i.e., light and radial velocity curves).

{\em iii)} Once the edges of the pulsation region in the HR diagram
are determined, the instability strip is populated according to a
period-mass distribution $P(M) \sim 1/M^3$ suggested by 
Kennicutt et al.\ (1998) and two different procedures for the 
Cepheid luminosity. For each mass and chemical composition, we adopted  
a) the canonical luminosity given by Eq. (1) (Caputo, Marconi, \& Musella 2000), 
and b) the evolutionary tracks computed by Pietrinferni et al.\ (2004) 
with a Reimers mass-loss parameter $\eta$=0.4 (see Fiorentino et al.\ 2007, 
for more details). In the latter case, we also take into account the 
evolutionary time spent by the Cepheids inside the strip.

{\em iv)} For each predicted pulsator, we calculate the period by 
means of the pulsation equation and the absolute magnitude in the 
various photometric bands by using the model atmospheres by Castelli,
Gratton \& Kurucz (1997a,b). Eventually, we derive with a standard 
regression the multiband $P$-$L$ relations. Note that these synthetic 
$P$-$L$ relations refer to {\it static} magnitudes, i.e., the magnitude 
the star would have, if it were not pulsating.

\begin{figure}
\begin{center}
\includegraphics[width=8cm]{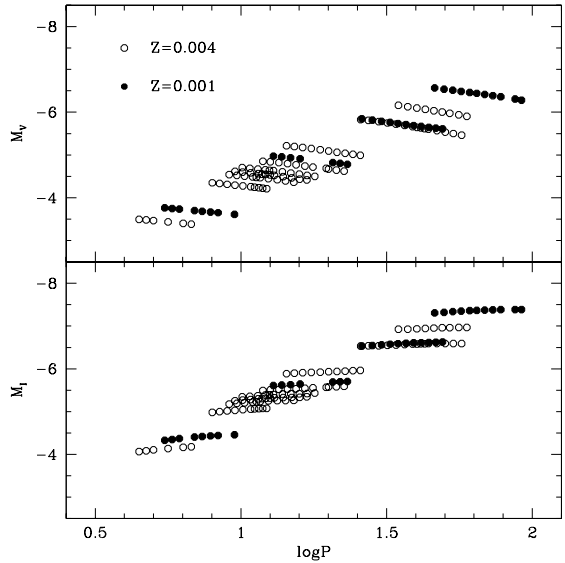}
\caption{Comparison between pulsation models with different metal contents, 
namely  $Z$=0.001 (solid circles) and $Z$=0.004 (open circles).}
\end{center}
\end{figure}

\begin{figure}
\begin{center}
\includegraphics[width=8cm]{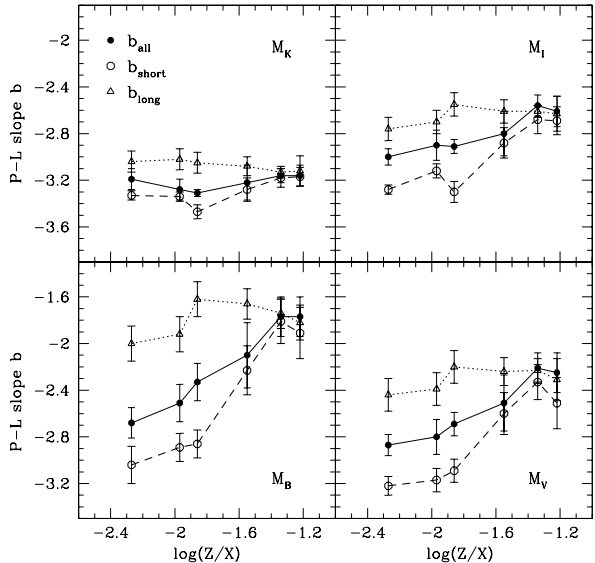}
\caption{Predicted slopes ($b$) of linear $P$-$L$ relations based on
pulsation models with metal content ranging from $Z$=0.004 to $Z$=-0.04. 
The standard regression including all the predicted pulsators gives the 
slopes $b_{all}$ (solid line), while the slopes $b_{short}$ (dashed line) 
and $b_{long}$ (dotted line) were evaluated using short- (log$P\le$1) and 
long-period (log$P>$1) pulsators, respectively.}
\end{center}
\end{figure}

As a whole, the synthetic $P$-$L$ relations depend on the adopted 
$ML$ relation, on the procedure adopted to populate the instability strip, 
on the helium content $Y$ (at fixed $l$/$H_p$ and $Z$, see  
Fiorentino et al.\ 2002; Marconi, Musella, \& Fiorentino 2005) 
and on the mixing-length parameter $l/H_p$ (at fixed $Z$ and $Y$, 
see Fiorentino et al.\ 2007). However, the quoted parameters and 
assumptions mainly affect the zero-point of the  $P$-$L$ relations, but 
the slope shows minimal changes. According to the above evidence,  
we used all the synthetic populations of fundamental models to derive 
the  average $P$-$L$ slope for the different bands listed in Table~2 
as a function of the chemical composition parameter log($Z/X)$, 
where $X$ is the hydrogen content ($X=1-Z-Y$). 
We fit the predicted fundamental pulsators with 0.4$\le$log$P\le$2.0 with 
a linear regression $M_i=a+b$log$P$ to determine the slope $b_{all}$. 
Then, the pulsators with log$P\le$1 and log$P>$1 were used to derive 
$b_{short}$ and $b_{long}$, respectively. The predicted dependence of 
the overall slopes on the chemical composition $\partial b_{all}/\partial$log$(Z/X)$, 
in the quoted metallicity range, are also listed in Table~2. 
Note that no synthetic $P$-$L$ relations is presently available 
for $Z<$ 0.004 but we show in Fig.~1 that pulsation models recently 
computed with $Z$=0.001 and $Y$=0.24 (Marconi et al.\ 2010, in preparation) 
appear in close agreement with the Period-Magnitude location of more metal-rich  
pulsators ($Z$=0.004). 
This finding seems to suggest that the  slope of the $P$-$L$ relation 
is marginally affected by the metal abundance for chemical compositions 
more metal-poor than log$(Z/X)=-2.27$.

In order to make the following comparison between theory and observations easier, 
we give in Table~3 the oxygen and iron abundances of the pulsation models 
which were derived by adopting scaled-solar chemical 
compositions and the solar chemical composition (log$(Z/X)_{\odot}$=$-1.78$, 
12+log(O/H)$_{\odot}$=8.66) from Asplund et al.\ (2004). 

The slopes of the predicted $P$-$L$ relations plotted in Fig.~2 
(see also Table~2) as a function of log($Z/X)$ show 
that they become steeper by increasing the filter wavelength, in 
agreement with well-known empirical results 
(see, e.g., Madore \& Freedman 1991). An increase in the metal content 
causes a flattening in the optical ($BVI$) $P$-$L$ relations, while the 
slopes of the short-period relations (dashed lines) are typically steeper 
than those for long-period ones (dotted lines). Again, the amplitude of 
these two effects decreases as the filter wavelength increases.
In the NIR $JK$-bands, the dependence of the slope, both 
on the period range and on the metal content, is significantly 
reduced. 

For a first comparison with observations, we list in Table~4 the 
slopes for the $P$-$L$ relations of fundamental Cepheids available 
in  the most recent literature for the Magellanic Clouds, the Milky Way 
(MW) and M31.

%%%%%%%%%%%%%%%%%%%%%%%%%%%%%%%%%%%%%%%%%%%%%%%%%%%%%%%%%%%%%%%%%%%%%%
\subsection{Magellanic Clouds}

\begin{figure}
\begin{center}
\includegraphics[width=8cm, height=0.45\textheight]{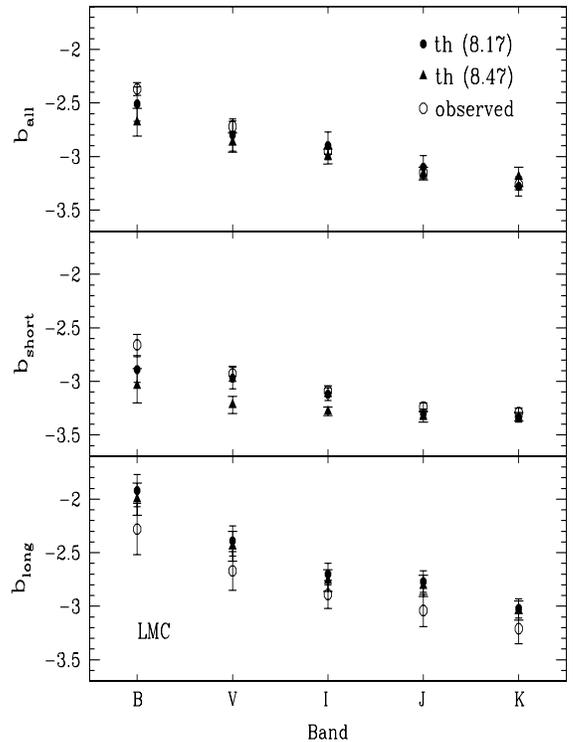}
\caption{Comparison between observed slopes of LMC $P$-$L$ relations 
(open circles) with the predicted values for a chemical composition 
of  12+log(O/H)=8.17 (filled circles) and of 8.47 (filled triangles).}
\end{center}
\end{figure}

The investigations focussed on LMC Cepheids, the most accurate 
sample, disclose the expected steepening of the $P$-$L$ relations, 
when moving from optical to NIR magnitudes. Moreover, they show  
that the optical $P$-$L$ relations of short (log$P\le$1) and long-period 
(log$P>$1) Cepheids do have statistically different slopes with the evidence 
of a flattening in the long period regime. On the other hand, the slopes 
of the $P$-$L_K$ relation do not show any significant difference 
(see the discussions in Sandage, Tammann, \& Reindl 2004, hereinafter STR04;
and Ngeow, Kanbur, \& Nanthakumar 2008, hereinafter NKN08). The quoted trends 
agree quite well with current predictions, and indeed data plotted in Fig.~3 show 
that the observed slopes attain {\it values} similar to the predicted 
ones for a chemical composition of 12+log(O/H)=8.17 and of 8.47 dex.

\begin{figure}
\begin{center}
\includegraphics[width=8cm, height=0.45\textheight]{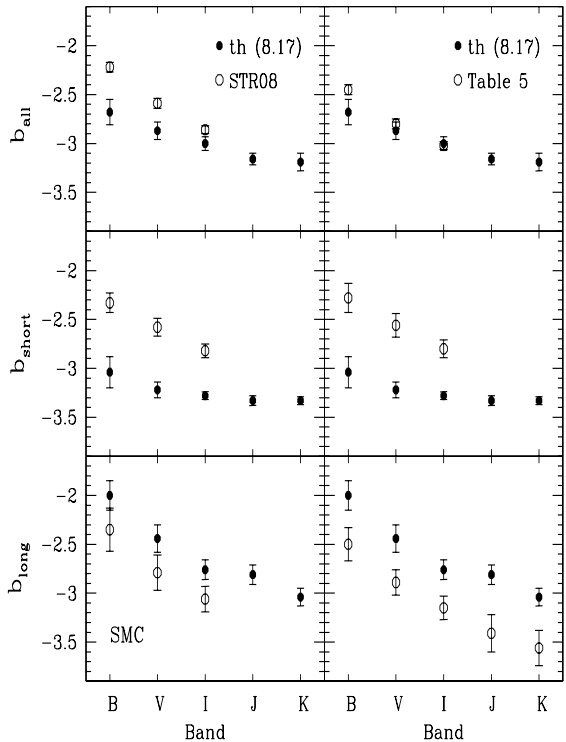}
\caption{Comparison between predicted slopes for a metal content of 
$Z$=0.004 (filled circles) and observed values of the SMC $P$-$L$ relations 
(open circles), using the results by STR08 (left panel) and the ones from 
our Table 5 (right panel).}
\end{center}
\end{figure}

The analysis by STR08 of the SMC variables relies on OGLE $BVI$
data. This sample includes 344 Cepheids with periods ranging from 
log$P\sim$0.50 to log$P\sim$1.62   
and seems to suggest a steepening in the long-period range, although 
the difference in the slope between short- and long-period Cepheids 
is within $\sim 1\sigma$ and less significant than for LMC Cepheids. 
In order to extend the analysis of SMC Cepheids to periods longer than 
the OGLE Cepheids and to near-infrared bands, we have used additional 
$BVI$ data (75 Cepheids, 0.93$\le$ log$P$ $\le$ 1.93) from 
Caldwell \& Coulson (1984) and $JK$ data (22 Cepheids, 
0.93$\le$ log$P$ $\le$ 1.93) from Laney \& Stobie (1986). 
Then, we fit with a standard regression 
all the variables with log$P\ge$0.5, as well as to the subsets of 
short- and long-period Cepheids. The results on the slopes are 
summarized in Table~5. 

By taking into account the slopes listed in Table~4 and in Table~5, 
we reach the conclusion that also the SMC Cepheids indicate a 
nonlinear $P$-$L$ relations.
In particular, data plotted in Fig.~4 display a reasonable agreement 
between the observed slopes $b_{all}$ and the predicted slopes for 
12+log(O/H)=8.17, whereas the observed $b_{short}$ values  
are systematically flatter, and the observed $b_{long}$ values 
are systematically steeper than current predictions.

%%%%%%%%%%%%%%%%%%%%%%%%%%%%%%%%%%%%%%%%%%%%%%%%%%%%%%%%%%%%%%%%%%%%%%
\subsection{Milky Way and M31}

The published slopes for the MW and the M31 variables were estimated 
over the entire period range. 
% Point B  
Here, we only wish to note that the Galactic slopes determined by Benedict et al. (2007,
hereinafter Be07) using trigonometric parallaxes, by Fouqu{\'e} et al.\ (2007,
hereinafter Fo07) using distances based on different methods and by Groenewegen (2008,
hereinafter Gr08),   
who adopted a revised projection factor 
$p$ to estimate the distances based on the Baade-Wesselink method, 
are significantly flatter than those found by STR04.
The slopes we found for the Galactic and the M31 fundamental Cepheids
are listed in Table~5. In the former case, we adopted  $BVI$ magnitudes 
from Berdnikov et al.\ (2000), $JK$ magnitudes from Berdnikov et al.\ (1996), 
reddenings and distances from Fo07 and Gr08 (74 Cepheids, 0.6$\le$ log$P$ $\le$ 1.8) 
% point A 
Note that the quoted authors use the extinctions provided by Laney \&
Caldwell (2007), the reddening law by Cardelli et al. (1989) and total to
selective absorption ratios of $R_V$=3.23 (Fo07) and of 3.3 (Gr08),
respectively. In the following we adopt $R_V$=3.23 ($A_V$/($A_V$-$A_I$)=2.55,
the same value adopted by the OGLE project).
For the M31 variables, we adopted the sample investigated by Tammann, 
Sandage, \& Reindl 2008, hereinafter TSR08), but 
the observed magnitudes were not unreddened with reddening corrections 
based on the Galactic Cepheids. We decided to follow this approach 
because, as noted by TSR08, the assumption that the M31 Cepheids 
obey to the Galactic Period-Color relation gives  
the unpleasant result of individual reddenings which increase with
the period. Note that the original near-infrared magnitudes of SMC and
MW Cepheids have been transformed into the 2MASS photometric system 
by using the transformations provided by Fo07.

\begin{figure}
\begin{center}
\includegraphics[width=8cm]{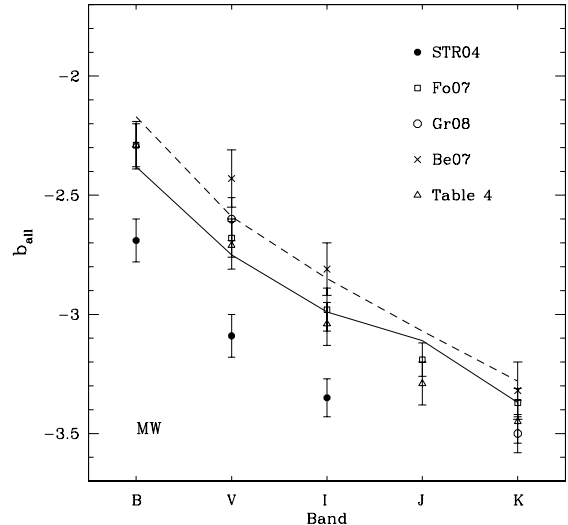}
\caption{Comparison between observed slopes of the Galactic
$P$-$L$ relations (open symbols) and the predicted values for a chemical 
composition of 12+log(O/H)=8.58 (solid line) and 12+log(O/H)=8.89 
(dashed line).}
\end{center}
\end{figure}

The occurrence of a nonlinear $P$-$L$ relation can be barely detected 
for MW and M31 Cepheids, likely due to the uncertainties on individual 
distances and on reddening corrections. However, it is worth noting 
that our overall $P$-$L$ relations are flatter 
than those determined by STR04 and TSR08, but our Galactic $b_{all}$ 
slopes are almost identical to those provided by Fo07 and Gr08. 
Data plotted in Fig.~5 show that all the $b_{all}$ values, apart from 
those determined by STR04, for Cepheids with a solar-like chemical composition 
are in reasonable agreement with predicted slopes for metal-rich chemical 
compositions, namely 12+log(O/H)=8.58 (solid line) and 
12+log(O/H)=8.89 (dashed line).

In conclusion, the predicted nonlinear feature of the optical $P$-$L$
relations seems verified by LMC and SMC Cepheids, but with opposite 
slope differences between short- and long-period variables. 
Moreover, we found that the observed {\it overall} slopes 
for the Magellanic, MW and M31 Cepheids agree quite well with the 
predicted values and suggest a flattening towards more metal-rich 
chemical compositions. This finding is at variance with the steepening 
found by TSR08, STR08, and by Sandage \& Tammann (2008, hereinafter ST08)
and it will be discussed in the next section.

%%%%%%%%%%%%%%%%%%%%%%%%%%%%%%%%%%%%%%%%%%%%%%%%%%%%%%%%%%%%%%%%%%%%%%%%%%%
\section{Dependence of the P-L slope on the Cepheid metallicity}

In Table~6, we list the observed $b(V)$ and $b(I)$ slopes of
single-fit $P$-$L$ relations according to TSR08 and to 
Saha et al.\ (2006, hereinafter STT06). 
\begin{figure}
\begin{center}
\includegraphics[width=8cm]{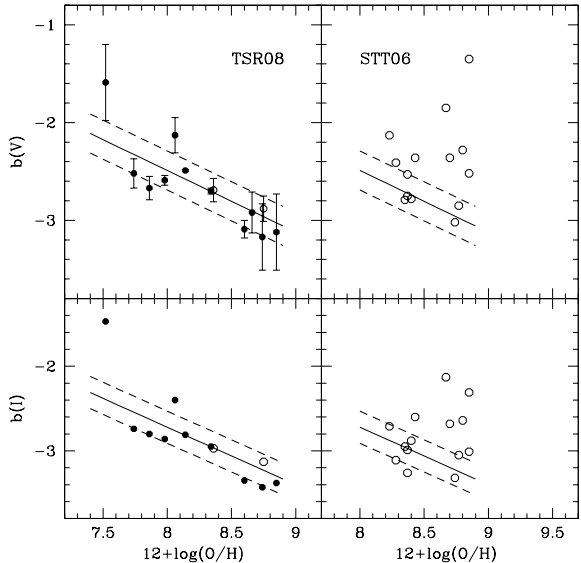}
\caption{Left -- Observed $b(V)$ and $b(I)$ slopes according to 
TSR08 as a function of the galaxy metal content. The open circles are 
the mean values for the metal-poor and the metal-rich galaxies according 
to STT06. The solid lines are the least square fits to the data, 
while the dashed lines display the 1$\sigma$ dispersion around the fit
(see text). The  couple Sextans (A+B) was not included in the fit. 
Right -- Observed $b(V)$ and $b(I)$ slopes for all the galaxies 
studied by STT06. Both solid and dashed lines have the same meaning 
as in the left panels.}
\end{center}
\end{figure}
The left panels of Fig.~6 show the TSR08 slopes (filled circles) versus 
the galaxy metallicity, together with the means (open circles) of 
metal-poor (M-P) and metal-rich (M-R) galaxies from STT06 according to 
TSR08. These data show an indisputable steepening of the $P$-$L$ relation 
with increasing galaxy metal abundance. We excluded the couple Sextans (A+B), 
which has the lowest oxygen abundance, since their $P$-$L$ relation are 
based on a few Cepheids ($N_{Ceph}$$\le$ 10), and we found  

$$b(V)=-2.70-0.63[\log(O/H)+3.66]\eqno(2)$$
$$b(I)=-2.95-0.68[\log(O/H)+3.66]\eqno(3)$$

\noindent 
with standard deviations around the fit of 
$\sigma_{b(V)}$=0.20 and $\sigma_{b(I)}$=0.19, respectively.
However, the right panels of Fig.~6 show quite clearly that 
such a trend can hardly explain the location on this plane of
several metal-rich galaxies studied by STT06, which actually 
present quite flat $P$-$L$ relations deviating more than 
2$\sigma$ from the TSR08 relations.

\begin{figure}
\begin{center}
\includegraphics[width=8cm]{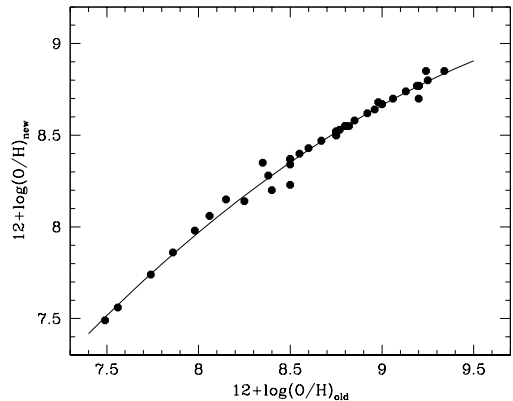}
\caption{Relation between new and old oxygen abundances given by 
Sakai et al.\ (2004) and by STT06. The solid line is the polynomial fit 
adopted in the current paper to derive the chemical compositions listed 
in Table~7.}
\end{center}
\end{figure}

To shed new light on this interesting evidence, we took into account 
all the galaxies with Cepheid photometry available in the recent 
literature and performed a new estimate of the slope of the 
$P$-$L$ relations. In particular, we adopted all the fundamental 
variables with 0.5$\le$log$P\le$2.0, the observed magnitudes 
were selected adopting a 2$\sigma$ clipping and we assumed that 
the reddening is independent of the Cepheid period. 
Concerning the metallicity of these extragalactic variables, 
we remind that they are generally based on the nebular oxygen abundance 
of the host galaxy and that the abundance scale provided by 
Zaritsky, Kennicutt \& Huchra (1994, hereinafter ZKH) has been recently 
revised by Sakai et al.\ (2004, hereinafter Sa04) and by STT06. 
Their data concerning {\it old} and {\it new} values are listed in 
Table~7 and are plotted in Fig.~7 together with the polynomial regression 
which has been used in the current paper to transform additional abundances 
in the ZKH scale into the new abundance scale.      

The full list of the observed $b(V)$ and $b(I)$ slopes derived in the 
present paper is given in columns (3) and (4) of Table~8, together with 
the galaxy metallicity in the new scale and the appropriate reference. 
Several galaxies were either observed or revised by the 
"Extragalactic Distance Scale Key Project" (hereinafter KP) 
or by the "Type Ia Supernova Calibration Project" (hereinafter SNP),  
or by other authors and we give the results for each galactic sample.

\begin{figure}
\begin{center}
\includegraphics[width=8cm]{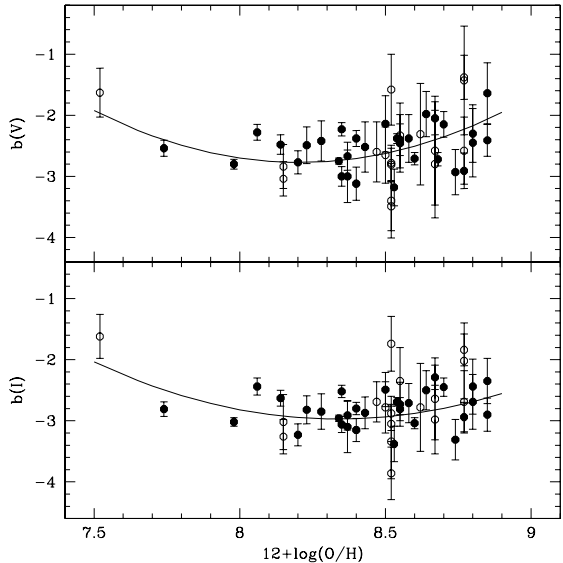}
\caption{Observed $b(V)$ and $b(I)$ slopes listed in Table~8, 
as a function of the galaxy metal content. 
Filled circles display slopes based on at least 20 Cepheids. 
The solid lines depict the quadratic fit to the data.
}
\end{center}
\end{figure}

Looking at the slopes listed in Table~8, we note that 
different photometric data for the same galaxy may lead to 
quite different slopes, in particular for the galaxies with 
few Cepheids. This is an unpleasant result, but the real 
gist of the matter is shown in Fig.~8, where all our $P$-$L$ 
slopes are plotted as a function of the galaxy metal content. 
In spite of the large spread of the $P$-$L$ slopes at fixed metal 
abundance, we are facing the evidence that all the Cepheids suggest 
a puzzling parabolic metallicity dependence with a sort of "turn-over" 
around 12+log(O/H)$\sim$ 8.2. The less metal abundant galaxies show 
a flattening of the $P$-$L$ relation with decreasing metallicity,
whereas the more metal abundant galaxies show an opposite 
trend.
% Point D
Note that the most metal-poor galaxy (12+log(O/H)$\sim$7.5) plotted in the above figure
is Sextans A (see also Table~8).

\begin{figure}
\begin{center}
\includegraphics[width=8cm]{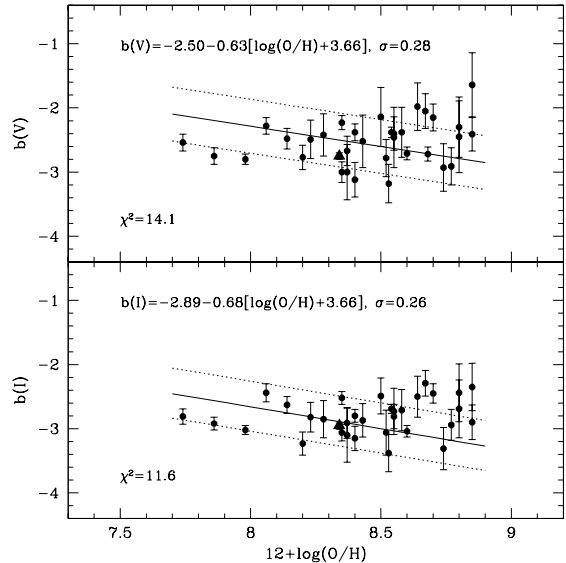}
\caption{Same as Fig.~8, but using only galaxies with at least 20 Cepheids. 
The solid lines are drawn according to the labeled relations, 
which adopt the metallicity dependence of Eq. (2) and Eq. (3), while 
the dotted lines depict the 1.5$\sigma$ dispersion around them. The 
$\chi^2$ values are derived by comparison of observed frequencies 
with those predicted by a normal distribution (see text).}  
\end{center}
\end{figure}

\begin{figure}
\begin{center}
\includegraphics[width=8cm]{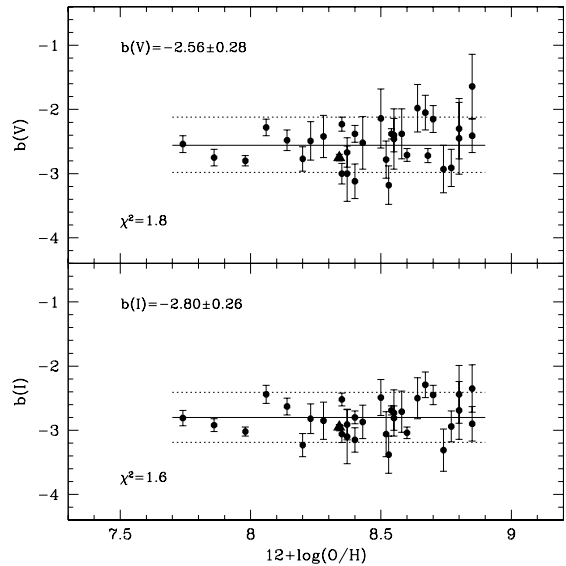}
\caption{Same as in Fig.~9, but assuming no dependence of the $b(V)$ and 
of the $b(I)$ slopes on the galaxy metal abundance.}
\end{center}
\end{figure}

\begin{figure}
\begin{center}
\includegraphics[width=8cm]{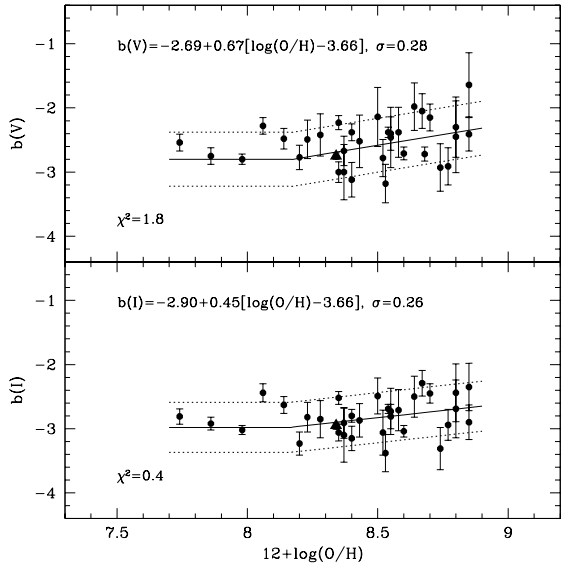}
\caption{Same as in Fig.~9, but assuming the metallicity dependence predicted 
by the pulsation models for chemical compositions with log$(Z/X)\ge-$2.27 
and a constant slope for log$(Z/X)<-$2.27.} 
\end{center}
\end{figure}

In order to overcome possible spurious effects due to limited samples, 
we only took into account galaxies with at least 20 Cepheids (filled 
circles in Fig.~8), and we compared the observed slopes with three 
selected metallicity dependences.
\begin{enumerate}

\item {\tt Steady steepening of the $P$-$L_{V,I}$ relations with 
increasing metal content.} 

By using the metallicity dependence of 
Eq. (2) and Eq. (3) and the ensuing zero-points of the $b(V)$ and 
of the $b(I)$ relations ($-2.50\pm$0.28 and $-2.89\pm$0.26), the 
slopes plotted in Fig.~9 show that several metal-rich galaxies 
deviate more than +1.5$\sigma$ (dotted line). 
The $\chi^2$ test on the observed frequencies in the intervals 
$<-$1.5$\sigma$, $-$1.5$\sigma \div$ 0, 0$\div$+1.5$\sigma$ and 
$>$+1.5$\sigma$ when compared with a normal distribution gives 
$\chi^2_{b(V)}$=14.1 and $\chi^2_{b(I)}$=11.6. These values,  
for 3 degrees of freedom, mean that the above hypothesis can be 
discarded with a confidence level of 99\%. 

\item {\tt Null hypothesis.} 

By assuming no dependence of the slopes 
on the metal content (see Fig.~10) and by using the mean values 
$b(V)=-2.56\pm0.28$ and $b(I)=-2.80\pm0.26$, the  $\chi^2$ test 
gives  $\chi^2_{b(V)}$=1.8 and $\chi^2_{b(I)}$=1.6. This means that 
we can accept the hypothesis with a confidence level of 62\% 
and of 67\%, respectively.

\item {\tt Trend predicted by pulsation models.} 

Fig.~11 shows that the slope of the $P$-$L_V$ and of the $P$-$L_I$ 
relation becomes shallower with increasing metal content for 12+log(O/H)$\ge$8.17 
and constant for lower metal abundances as suggested by the pulsation models. 
In this case, the  $\chi^2$ test gives $\chi^2_{b(V)}$=1.8 and 
$\chi^2_{b(I)}$=0.6, i.e. the theoretical hypothesis can be accepted with 
a confidence level of 62\% and 92\%, respectively.
\end{enumerate}

In conclusion, the slopes of the observed $P$-$L_V$ and $P$-$L_I$ relations 
seem to exclude the steepening with increasing metallicity suggested by 
TSR08. Current findings suggest that the slopes are either metallicity 
independent or they follow the metallicity dependence predicted by the 
pulsation models. In the latter case, the use of these $P$-$L$ relations 
for distance determinations {\it could} require specific corrections to  
account for the difference in the slope. This issue is discussed in the 
next section, where we show the impact of the $P$-$L$ slopes on the 
Cepheid distance scale.

%%%%%%%%%%%%%%%%%%%%%%%%%%%%%%%%%%%%%%%%%%%%%%%%%%%%%%%%%%%%%%%%%%%%%%%%%%%%%%%
\section{Intrinsic features of the P-W relations}

The fiducial $P$-$L$ relations based on the unreddened magnitudes of 
the LMC Cepheids can be used to derive the LMC-relative apparent 
distance modulus $\delta\mu_i$ of a given variable and the use 
of two or more passbands provides the opportunity to estimate the 
reddening, e.g., $\delta\mu_B-\delta\mu_V=A_B-A_V=E(B-V)$. 
Then, the use of a wavelength-dependent extinction law yields the 
correction for the selective extinction in the different bands, and 
eventually the LMC-relative true distance modulus $\delta\mu_0$.

This approach is equivalent to the method based on the so-called
Wesenheit functions; since the early papers on Cepheid
investigations (see Madore 1982; Madore \& Freedman 1991) the
problem of dust extinction is generally accounted for using the
Cepheid colors to derive extinction-free magnitudes --e.g., 
$WBV=V-R_V(B-V)$, where $R_V$ is the visual extinction-to-reddening 
ratio $A_V/E(B-V)$-- and to use them in Period-Wesenheit ($P$-$W$) relations 
that are independent of reddening. We note that the effect of 
the extinction is similar to the one produced by the finite width of the 
instability strip, therefore, the scatter around the $P$-$W$ relations 
is smaller than in any observed $P$-$L$ relation. Thus, we are facing 
the circumstantial evidence that {\it the dependence on metallicity 
of the slope of a monochromatic $P$-$L$ relation could have a minimal 
impact upon the distance scale if the combination of different magnitudes  
and colors leads to a metallicity independent slope for the 
$P$-$W$ relations}. 

On the theoretical side, the bolometric light curves provided by 
the nonlinear approach, once transformed into the observational plane  
by using model atmospheres, provide the amplitudes in the various spectral 
bands and, after a time integration, the predicted mean magnitudes of 
the pulsators. In Table~9, we give the predicted slope $\beta(W)$  
of selected $P$-$W$ relations ($W=\alpha+\beta$log$P$) for fundamental 
pulsators with 0.5$\le$log$P\le$2.0, based on intensity-weighted mean 
magnitudes $\langle M_i\rangle$ and colors [$\langle M_i\rangle -\langle M_j\rangle$] 
and on the Cardelli et al.\ (1989) reddening law with $R_V$=3.23. Note that the 
original near-infrared magnitudes of our models, which are in the 
Bessell \& Brett photometric system, have been transformed into the 
2MASS system using the conversion relations adopted by Fo07. 

The predicted $P$-$W$ relations are quite tight, 
and show several additional undeniable advantages: 

%\begin{itemize}
{\em i)} they are almost independent of the pulsator distribution within 
the instability strip. For this reason, in Table~9 we also give the 
results for the $Z$=0.001 models; 

{\em ii)}  they are almost linear over the entire period range; 

{\em iii)}  the effects of the mixing-length --$l$/$H_p$-- at fixed metal 
--$Z$-- and helium --$Y$-- content on both the slope and zero-point 
are negligible;

{\em iv)}  the adopted helium content $Y$, at fixed metal content $Z$ and 
mass-luminosity relation, only affects the zero-point of the relations; 

{\em v)}  their slopes turn out to be almost independent of the chemical 
composition. 
%\end{itemize} 

The above issues have been discussed in our previous
papers (see Fiorentino et al.\ 2007; Bono et al.\ 2008, and references 
therein), where we showed that the impact of the metallicity on the 
$P$-$W$ zero-point also depends on the adopted Wesenheit function.
Here, we test these predictions by directly using Cepheid observations.

\begin{figure}
\begin{center}
\includegraphics[width=8cm]{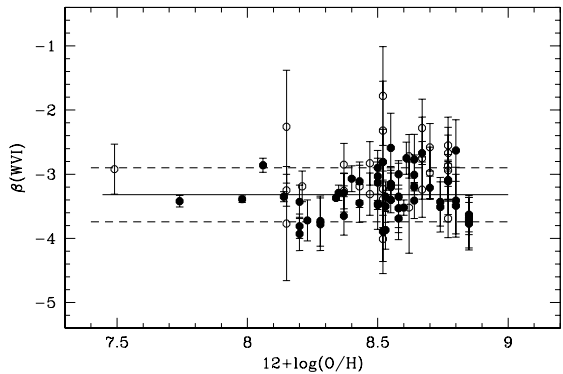}
\caption{Slopes of the observed $P$-$WVI$ relations for the same galaxies 
plotted in Fig.~8. The solid and dashed lines refer to the mean value and 
to the 1.5$\sigma$ dispersion, respectively.}
\end{center}
\end{figure}

Firstly, the linearity of the predicted $P$-$W$ relations agrees 
with the recent results by Ngeow \& Kanbur (2005) and by Ngeow et al.\ (2005)
from measurements of LMC Cepheids and, as recently discussed by 
Madore \& Freedman (2009), is expected even though 
{\it the underlying monochromatic  $P$-$L$ relations are nonlinear}. 
Secondly, the slope of our own LMC relations

$$WBV=16.04(\pm0.04)-3.82(\pm0.06)\log P, (\sigma=0.19)\eqno(4)$$
$$WVI=15.91(\pm0.03)-3.37(\pm0.03)\log P, (\sigma=0.09)\eqno(5)$$
$$WJK_s=15.97(\pm0.07)-3.45(\pm0.06)\log P, (\sigma=0.10)\eqno(6)$$

\noindent 
are quite consistent with the predicted values listed in Table~9.   
Finally, previous studies have already shown that the slope of the 
$P$-$W$ relations defined by 
% Point B 
Cepheids in the Galaxy (Be07; van Leeuwen et al. 2007), in metal-poor 
galaxies (Pietrzynski et al. 2007) and in metal-rich ones (Riess et al.\ 2009, 
hereinafter Ri09) agrees quite well with the slope of the LMC variables.
This result is supported by the $\beta(WVI)$ 
values derived in the present paper, which are listed in the last 
column of Table~8 and are plotted versus the galaxy oxygen abundance 
in Fig.~12. In particular, we find that only the slopes based on 
small Cepheid samples ($N_{Cep}<$20, open circles) deviate more than 2$\sigma$ 
from the weighted mean $\beta(WVI)=-3.32\pm0.21$, which in turn 
is practically identical to the LMC value.     
Moreover, the new slopes of the observed $P$-$WBV$ and $P$-$WJK_s$ 
relations listed in Table~10 are also very similar to the LMC values. 
There are only two exceptions, namely Sextans~A and the outer region 
of NGC~598. 
 
In conclusion, empirical findings strongly support the use of the 
$P$-$W$ relations based on LMC Cepheids to derive the LMC-relative 
true distance modulus of individual Cepheids, 
{\it provided that the metallicity dependence of the zero-point of 
the different $P$-$W$ relations, if present, is taken into account.} 

We recall that independent observations suggest either a negligible 
metallicity effect on the Cepheid distance scale or that variables 
in metal-rich galaxies are, at fixed period, {\it brighter} than 
those in metal-poor galaxies (see, e.g., Sasselov et al.\ 1997; 
Kennicutt et al.\ 1998,2003; Kanbur et al.\ 2003; Tammann et al.\ 2003; 
Sandage et al.\ 2004; Storm et al.\ 2004; Groenewegen et al.\ 2004; 
Sakai et al.\ 2004; Ngeow \& Kanbur 2004; Pietrzynski et al.\ 2007). 
In the latter case, the various estimates of the parameter 
$\gamma=c/\delta$log$Z$, where $c$ is the size (in magnitude) 
of the metallicity correction and $\delta$log$Z$=log$Z_{LMC}-$log$Z_{Ceph}$, 
depends on the wavelength  but always give {\it negative} numbers, 
with an average value of approximately $-$0.27~mag~dex$^{-1}$ 
(see the discussion in Groenewegen 2008). However, recent spectroscopic
iron-to-hydrogen [Fe/H] measurements of Galactic and Magellanic Cepheids 
(Romaniello et al.\ 2005,2008; Groenewegen 2008) seem to suggest that 
the $P$-$L_V$ relation becomes {\it fainter} with increasing metallicity. 
Current data provided no firm conclusion concerning 
the dependence of $P$-$L_K$ relation on the metal content. 
 
By using the pulsation models listed in Table~1, we can derive, by assuming 
that these are actual Cepheids located at the same distance ($\mu_0$=0) 
and with different chemical compositions, from equations (4)-(6) their 
LMC-relative true distance moduli $\delta\mu_0(WBV)$, 
$\delta\mu_0(WVI)$ and $\delta\mu_0(WJK_s)$. By averaging the results 
over the period range 0.5$\le$log$P\le$2.0, we derived the values at 
log$L/L_{can}$=0 and 0.2  (see Table~11). They clearly show that the 
$\delta\mu_0$ values mainly depend on the filter and on the 
$L/L_{can}$ ratio. Moreover, the metallicity effect on $\delta\mu_0(WBV)$ 
is quite strong and at constant $L/L_{can}$ ratio we found 
$\gamma(WBV)=-0.58(\pm0.03)$~mag~dex$^{-1}$, where the error 
takes into account the different helium abundances at constant $Z$. 
On the other hand, the variations of $\delta\mu_0(WVI)$ and of 
$\delta\mu_0(WJK_s)$ are significantly smaller and both the extent 
and the sign of the metallicity effect seem to depend both on 
the helium content (at fixed $Z$), and on the adopted metallicity 
range (see also \S 5). 
  
It is worth underlining that the {\it internal differences} 
among the various LMC-relative distance moduli listed in Table~12 depend 
on the metal content, but they are independent of the helium content 
(at fixed $Z$) and of the adopted $ML$ relation. Specifically, the 
differences:  
$\Delta(WBV-WVI)=\delta\mu_0(WBV)-\delta\mu_0(WVI)$ and 
$\Delta(WBV-WJK_s)=\delta\mu_0(WBV)-\delta\mu_0(WJK_s)$ are quite 
sensitive to the metal content, and could provide a robust diagnostic to 
estimate the Cepheid metal content, since they are independent of uncertainties 
affecting reddening corrections.

On the observational side, let us use once again Eq.s (4)-(6) to 
derive the LMC-relative Cepheid distance moduli in external galaxies 
for which are available $BVIJK_s$ data. 
We show in Table~13 that the $P$-$WVI$ and $P$-$WJK_s$ relations 
provide similar results, whereas the $P$-$WBV$ relation, for galaxies 
more metal poor than the LMC, yields larger distances. As a whole, we 
found $\partial\Delta(WBV-WVI)/\partial$log(O/H)$\sim-$0.4~mag~dex$^{-1}$ 
which is in fair agreement with the value $-0.5$~mag~dex$^{-1}$ inferred 
from the predicted differences listed in Table~12.

\begin{figure}
\begin{center}
\includegraphics[width=8.5cm, height=12.0cm]{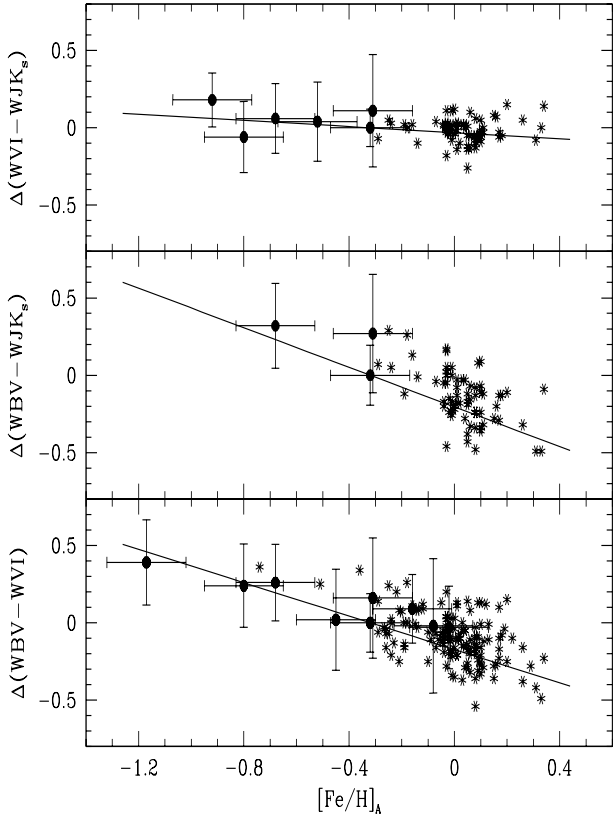}
\caption{Metallicity dependence of the differences among LMC-relative 
distance moduli for the external galaxies listed in Table~13 and for 
the Milky Way Cepheids with log$P>$0.5. For the external galaxies, we 
adopted [Fe/H]=log(O/H)+3.34 with a typical error of $\pm$0.15 dex, 
while for the Galactic Cepheids we adopted the iron measurements provided 
by Andrievsky and collaborators ([Fe/H]$_A$). The solid line shows the 
predicted trend according to the differences listed in Table~12.
}
\end{center}
\end{figure}

\begin{figure}
\begin{center}
\includegraphics[width=8.5cm,  height=12.0cm]{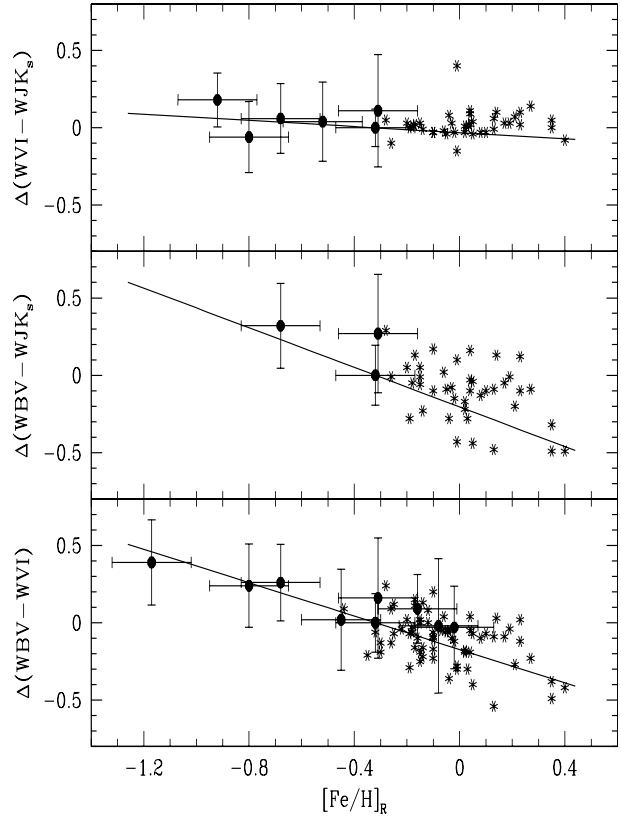}
\caption{Same as in Fig.~13, but with [Fe/H]$_R$ values by Romaniello and 
collaborators. 
}
\end{center}
\end{figure}

To provide an independent test and to include in the current analysis  
more metal-rich Cepheids, we applied the above procedure to the Galactic 
Cepheids with measured iron-to-hydrogen ratios, without the need to know 
their distance. 
The source of the $BVIJK_s$ magnitudes was already mentioned in \S 2 and, 
to overcome possible systematic uncertainties in abundance 
measurements, we adopted the two largest sets of Cepheid metallicities  
available in the literature: the iron abundances [Fe/H]$_A$ 
provided by Andrievsky and collaborators
(Andrievsky et al.\ 2002a,b,c; Andrievsky et al.\ 2004; Luck et al.\ 2003;
Kovtyukh, Wallerstein \& Andrievsky 2005; Luck, Kovtyukh \& Andrievsky 2006) 
and the iron abundances [Fe/H]$_R$ by Romaniello et al.\ (2005, 2008) and 
Lemasle et al.\ (2007,2008). We selected the Galactic variables with 
log$P>$0.5, although the inclusion of first overtone pulsators has no 
significant impact upon the {\it differences} among the $\delta\mu_0(W)$ 
values.

Eventually, Figures 13 and 14 show that extragalactic and Galactic
Cepheids follow well defined common relations which are fully consistent 
with the predicted behaviors presented in Table~12, and further confirm 
that the metallicity effect on the $P$-$W$ relations significantly 
depends on the adopted Wesenheit function. 

As a consequence, if the LMC-based $P$-$W$ relations are used to estimate 
the distance to Cepheids with metal content significantly different from 
the LMC abundance, then the values of the different relative distances  
$\delta\mu_0(W)$ {\it should not be averaged, but taken into account 
individually, to further exploit the information provided by their different 
metallicity dependence.}

%%%%%%%%%%%%%%%%%%%%%%%%%%%%%%%%%%%%%%%%%%%%%%%%%%%%%%%%%%%%%%%%%%%%%%%%%%%%
\section{Metallicity correction(s) to the Cepheid distance}

A straightforward estimate of the metallicity effect on the Cepheid distance 
scale can be determined by using regions of the same galaxy located at different 
galactocentric distances and characterized by significantly different metal 
abundances. After the pioneering work by Freedman \& Madore (1990) in M31, 
such a differential test was performed in NGC~5457 (Kennicutt et al.\ 1998, 
$\gamma$=$-0.24\pm$0.16~mag~dex$^{-1}$), in NGC~4258 (Macri et al.\ 2006, 
$\gamma$=$-0.29\pm$0.11~mag~dex$^{-1}$), and in NGC~598 (Scowcroft et al.\ 2009, 
hereinafter Sc09, $\gamma$=$-0.29\pm$0.11~mag~dex$^{-1}$) giving an average 
value of $\gamma$=$-0.27$~mag~dex$^{-1}$. 

However, it is worth mentioning that the metallicity gradient in NGC~598, 
as already discussed by Sc09, is still matter of debate and the use of the 
abundances of HII regions provided by Crockett et al.\ (2006) would 
provide a metallicity coefficient that is one order of magnitude larger 
($\gamma=-2.90$~mag~dex$^{-1}$). Moreover, the above results rely on the old metallicity 
scale provided by ZKH, and the use of the new metal abundances would imply 
significantly larger values of the metallicity coefficient $\gamma$  
with dramatic effects on the Cepheid distance scale. 
% Point C 
Moreover, Macri et al.\ (2001) showed that Cepheids in the inner field of
NGC~5457 are severely affected by blending, which gives artificially shorter
distances (Macri et al. 2006). Finally, we mention that van Leeuven et al. (2007),
using a $P$-$WVI$ relation based on Galactic Cepheids and Cepheids in the inner
field of NGC~4258, found a true distance (29.22$\pm$0.03 mag) that agrees quite
well with the maser distance (Herrnstein et al. 1999, 29.29$\pm$0.15 mag).
However, Mager et al.\ (2008) using Cepheids in the inner and in the outer field
of NGC~4258 suggested that the difference in the true distance modulus of the
two fields (29.26$\pm$0.03 vs 29.45$\pm$0.08 mag) might be due either to a
difference in the reddening law  or to small number statistics (see also
Di Benedetto 2008).
On these grounds, the quoted available differential tests in external 
galaxies need to be treated cautiously. 

Robust constraints on the dependence of Cepheid distances on metallicity 
can also be provided by the comparison with distances based on an 
independent distance indicator. 
The Tip of the Red Giant Branch (TRGB) appears to be a robust standard 
candle, largely independent of metallicity in the $I$-band 
(Madore, Mager, \& Freedman 2009; Sanna et al.\ 2009). 
%Point C 
However, we note that TRGB distances to galaxies characterized by complex star formation
histories and age-metallicity relations might also be biased ($\Delta$$\mu$$\le$0.10 mag) due
to the presence of TRGB stars younger than typical globular cluster counterparts
(Salaris \& Girardi 2005).
In the following, we shall refer to the TRGB distances given by 
Rizzi et al.\ (2007, hereinafter Ri07) and listed in column 2 
of Table~14.

\begin{figure}
\begin{center}
\includegraphics[width=9cm]{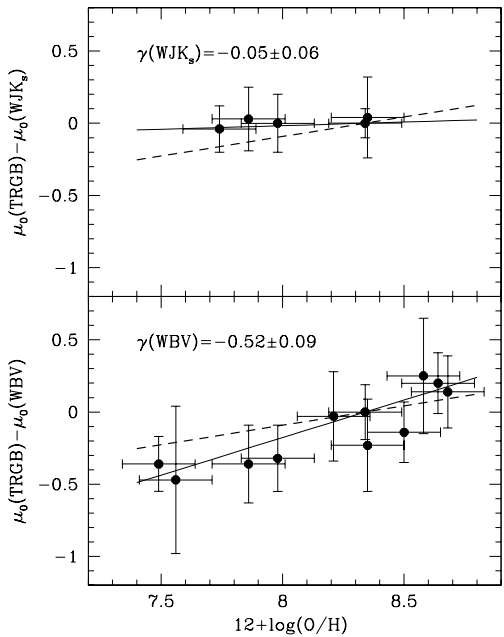}
\caption{Difference between TRGB and Cepheid distances based on the 
$P$-$WBV$ and on the $P$-$WJK_s$ relation as a function of the nebular 
oxygen abundance. The solid line is the least-square fit to the data, 
while the dashed line is the average empirical value of the metallicity 
coefficient $\gamma$=$-$0.27~mag~dex$^{-1}$.  
}
\end{center}
\end{figure}

In the previous section, we showed that the pulsation models predict  
a rather strong metallicity effect on the distance modulus determined 
by using the 
$P$-$WBV$ relation, whereas the distances based on the $P$-$WVI$ and 
$P$-$WJK_s$ relations appear rather independent of the Cepheid metal content. 
Data plotted in Fig.~15 clearly show that such a prediction is fully consistent 
with the observations, and indeed, the comparison between the measured 
$\delta\mu_0(WBV)$ and $\delta\mu_0(WJK_s)$ listed in Table~13 with the 
TRGB distances yields metallicity coefficients $\gamma$ that depend on 
the adopted passband. In particular, the TRGB distances provided by Ri07 give 
$-0.52\pm0.09$~mag~dex$^{-1}$ and $-0.05\pm0.06$~mag~dex$^{-1}$, respectively.
We note that these metallicity corrections agree quite well with the predicted 
values. 

The previously published Cepheid distances based on the $P$-$WVI$ relation 
are summarized in Table~15.  For the galaxies included in Table~8, the 
Cepheid sample  is listed in column (2), while column (3) gives the 
distance modulus published in the original paper, but scaled to a 
homogeneous LMC distance modulus of $\mu_0$=18.50~mag. 
The galaxy distances provided by Freedman et al.\ (2001, hereinafter Fr01) are listed 
in column (4). These distances were determined by using the LMC linear $P$-$L$ relations 
from Udalski et al.\ (1999a) and $\mu_0$(LMC)=18.50~mag. In column (5), we list 
the galaxy distances provided by Kanbur et al.\ (2003, hereinafter Ka03) using 
slightly revised LMC linear $P$-$L$ relations of the same data given by 
Udalski et al.\ (1999a) and $\mu_0$(LMC)=18.50~mag. Column (6) gives the 
distance moduli determined by STT06 using LMC $P$-$L$ relations with a 
break in the slope at $P$=10 days. Since STT06 adopt $\mu_0$(LMC)=18.54~mag, 
the original values were scaled to a distance modulus of 18.50~mag. 

A quick inspection of these data indicates that {\it on average} they are consistent 
with each other. However, for some individual galaxies, the difference among the various 
distances becomes of the order of $\pm$0.2-0.3~mag. Since this discrepancy might be the  
consequence of different assumptions and procedures used to derive distances, 
we decided to provide new homogeneous estimates of the $\mu_0(WVI)$ distance moduli for 
the entire sample. The individual distances are listed in column (6) by adopting the 
LMC relation of Eq.(5) together with $\mu_0$(LMC)=18.50~mag and by applying a $2\sigma$ 
clipping. 

We find that our distances are fully consistent 
with the values in the literature. The average difference 
between the new estimates and the distances listed in columns (3)--(6) 
are the following: $-$0.01$\pm$0.05~mag, +0.02$\pm$0.06~mag, 
+0.03$\pm$0.05~mag and +0.02$\pm$0.07~mag, respectively. 
However, even though we used the same procedure, individual galaxy distance 
moduli might show a difference of the order of $\sim$0.2-0.3~mag, according 
to the various Cepheid samples.   

\begin{figure}
\begin{center}
\includegraphics[width=9cm]{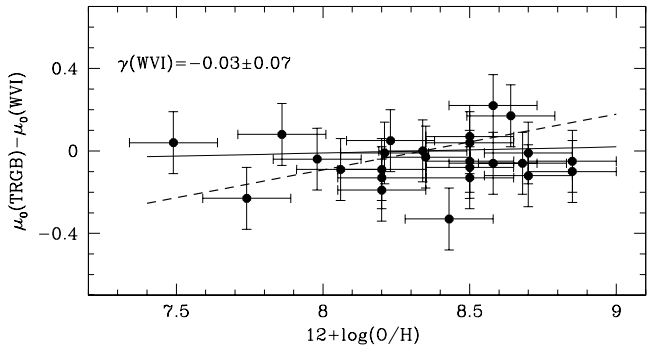}
\caption{Differences between TRGB and Cepheid distances based on the 
$P$-$WVI$ relation, as a function of the nebular oxygen abundance. 
The dashed line shows the average empirical value of the metallicity 
coefficient $\gamma=-$0.27~mag~dex$^{-1}$.  
}
\end{center}
\end{figure}

By taking into account the entire sample, we show in Fig.~16  
the comparison between galactic Cepheid distances based on the 
the $P$-$WVI$ relation and TRGB distances from Ri07. 
Once again, we found that the difference between Cepheid and 
TRGB distances give a vanishing metallicity correction 
--$\gamma(WVI)$=$-0.03\pm$0.07 mag dex$^{-1}$-- that agrees 
quite well with the predicted trend.

%%%%%%%%%%%%%%%%%%%%%%%%%%%%%%%%%%%%%%%%%%%%%%%%%%%%%%%%%%%%%%%%%%%%%%%%%%%%%
\section{Summary}

We performed a comprehensive investigation of the Cepheid distance scale 
by taking into account both theory and observations. In particular, we 
addressed the intrinsic features of both optical and NIR $P$-$L$ relation. 
Here are results: 

{\em i)} {\em Filter wavelength} Theory and observations indicate that 
the slopes of the $P$-$L$ relation become steeper when moving from optical 
to NIR bands.    

{\em iii)} {\em Nonlinearity} The slopes of the observed optical $P$-$L$ relations 
of Magellanic Cepheids are nonlinear. No firm conclusion was reached concerning the 
nonlinearity of the $P$-$L$ relations based on Galactic and M31 Cepheids.      

{\em ii)} {\em Period range} The slopes of NIR $P$-$L$ relations are less 
sensitive to the period range covered by Cepheids than the slopes of optical 
$P$-$L$ relation.  

{\em iv)} {\em Metal content} The derivative 
$\partial b_{all}/\partial$log$(Z/X)$ of the predicted slopes 
covering the entire period range decreases by more than a 
factor of two when moving from the $V$ to the $J$-band and by almost one order of 
magnitude when moving from the $V$ to the $K$-band. Moreover, the observed slopes   
of Magellanic, Galactic and M31 Cepheids agree quite well with predicted ones. In 
particular, they suggest a flattening of the slope when moving from metal-poor to 
metal-rich Cepheids. This finding is at odds with the steepening recently suggested 
by TSR08, by STR08 and by ST08.   

In order to provide an empirical estimate of the dependence of the  $P$-$L$ relation 
on metal content, we also adopted Cepheids in external Galaxies. To avoid possible 
deceptive uncertainties in the adopted metallicity scale, we derived a new relation  
to transform the old nebular oxygen abundances given by  Zaritsky et al.\ (1994) into 
the new metallicity scale provided by Sakai et al.\ (2004). 

Moreover, we provided new homogeneous estimates of  
$V$- and $I$-band slope for 87 independent Cepheid data sets available in the 
literature and 57 of them include more than 20 Cepheids. They are hosted 
in 48 external galaxies and for 27 of them two or more independent data sets 
are available. Four galaxies with multiple data sets (NGC~598, NGC~3031, NGC~4258, NGC~5457) 
have Cepheids located in an inner and in an outer galactic field. The galaxies with
more than 20 Cepheids cover a wide metallicity range (12+log(O/H)$\sim$7.7 [WLM],  
12+log(O/H)$\sim$8.9 dex [NGC~3351, NGC~4548])     
Note that the quoted range is approximately a factor of five larger than the 
metallicity range covered by SMC (12+log(O/H)$\sim$8) and Galactic 
(12+log(O/H)$\sim$8.6) Cepheids. By using Cepheid data sets larger than 20,  
we tested three hypotheses concerning the dependence of the $P$-$L$ relation 
on metal content:

{\em a)} {\em Correlation between the slope of the $P$-$L_{V,I}$ relations and 
the metallicity.}  The $\chi^2$ test on $V$- and $I$-band slopes indicates  
that this hypothesis can be discarded at the 99\% confidence level.  

{\em b)} {\em No dependence of the $P$-$L_{V,I}$ relations on the metallicity.}          
The $\chi^2$ test on $V$- and $I$-band slopes indicates that this hypothesis 
can be accepted, but only at the 62\% and 67\% confidence level. 

{\em c)} {\em Pulsation models predict that the slope of the $P$-$L_{V,I}$ relations 
becomes shallower in the metal-rich regime and constant in the metal-poor regime.} 
The outcome of the $\chi^2$ test on observed $V$- and $I$-band slopes is that 
the predicted trend can be accepted at the 62\% and 92\% confidence level.  

The main result of the above analysis based on external galaxies with sizable 
Cepheid samples is that the observed slopes of the  $P$-$L_I$ relation show the 
same metallicity trend predicted by pulsation models, while the slopes of the 
$PL_V$ relation either follow theory or do not depend on metallicity.  

Together with the $P$-$L_{V,I}$ relations we also investigated the reddening 
independent $P$-$W$ relations and the results are the following:    

{\em i)} {\em Dependence of the slope of the $P$-$W$ relations on metal content} 
Empirical estimates indicate that the slopes of optical ($P$-$WBV$,$P$-$WVI$) 
and NIR ($P$-$WJK_s$) relations in metal-poor and in metal-rich 
galaxies agree quite well with the slope of LMC Cepheids. This finding 
supports previous results by 
% Point B 
Benedict et al. (2007), Pietrzynski et al.\ (2007), van Leeuwen et al. (2007), 
and by Riess et al.\ (2009).
Moreover, it brings forward the evidence that 
the $P$-$W$ relations provide accurate estimates of LMC-relative true 
distance moduli. However, the metallicity dependence of the zero-point 
of the $P$-$W$ relations, if present, has to be taken into account.    
  
{\em ii)} {\em Use of the $P$-$W$ relations as a metallicity diagnostic} 
Current predictions indicate that LMC-relative true distance moduli 
based on the $P$-$WBV$ relation strongly depend on the metal content, 
whereas those based on the $P$-$WVI$ and on the $P$-$WJK_s$ relation 
minimally depend on metallicity. The difference between the quoted 
distances can provide estimates of individual Cepheid metallicities.         

The above findings further support the evidence that distances based 
on different $P$-$W$ relations should not be averaged, since the 
metallicity effect strongly depends on the adopted bands. 

Furthermore, we adopted the true distance moduli based on the TRGB method (Ri07) 
to validate the predicted metallicity corrections of the Cepheid distance scale. 
We found that the metallicity correction --$\gamma$-- obtained
using the TRGB distances agree quite well with pulsation predictions,
namely $\gamma(WBV)$=$-0.52\pm$0.09 mag dex$^{-1}$,
$\gamma(WVI)$=$-0.03\pm$0.07 mag dex$^{-1}$ and 
$\gamma(WJK)$=$-0.05\pm$0.06 mag dex$^{-1}$.

%%%%%%%%%%%%%%%%%%%%%%%%%%%%%%%%%%%%%%%%%%%%%%%%%%%%%%%%%%%%%%%%%%%%%%%%%%%%%
\section{Final remarks and future perspectives}

The results of this investigation rely on Cepheid samples which will soon 
become the templates to estimate the distances to late-type and 
dwarf irregular galaxies in the Local Group (d$\le$ 1 Mpc) and in the Local 
Volume (d$\le$ 10 Mpc). However, there are a few key points that demand future 
attention. 

{\em i)} NIR mean magnitudes are only available for a handful of systems. 
Cepheid distances based on the predicted $P-WVI$ and $P-WJK_s$ relations 
are minimally dependent on the metallicity. Therefore, the difference 
can provide firm constraints on the plausibility of the assumption of 
an universal reddening law. This gap will be certainly filled by the 
James Webb Space Telescope\footnote{More details can be found at the 
following URL: http://www.stsci.edu/jwst/} (Rieke et al.\ 2005) and 
the next generation of Extremely Large 
Telescopes\footnote{More details can be found at the following URL: 
http://www.eso.org/projects/e-elt/ and http://www.tmt.org/} 
(Gilmozzi \& Spyromilio 2009) will certainly fill.  

{\em ii)} $B$-band mean magnitudes are only available for a handful of 
systems. The difference between Cepheid distances based on the predicted 
$P-WBV$ and on the $P-WVI$/$P-WJK_s$ relations is a robust diagnostic 
to estimate the Cepheid metal content. The significant sensitivity in the 
$B$-band of the Wide Field Planetary 
Camera 3\footnote{More details can be found at the following URL: 
http://www.stsci.edu/hst/wfc3} (Wong et al.\ 2010) on board of the 
Hubble Space Telescope can play a relevant role in this context.           

{\em iii)} The quest for solid constraints on the precision of the Cepheid 
distance scale brought forward the need of an accurate and homogeneous 
metallicity scale for external galaxies. Current findings rely on two 
relevant assumptions. {\em a)} Oxygen abundances based on emission and on 
absorption lines appear to be rather similar (Hill 2004; 
Kudritzki et al.\ 2008; Bresolin et al.\ 2009), but new quantitative 
constraints in external galaxies are requested.            
{\em b)} The oxygen appears to be a good proxy of the iron content 
(STT06). But oxygen is an $\alpha$-element and it is not clear whether 
this approximation is still valid over the entire metallicity range.  

{\em iv)} The pulsation models currently adopted to constrain the properties 
of Galactic and external Cepheids were constructed assuming scaled-solar heavy 
element abundances. However, we still lack firm theoretical constraints on the 
impact of $\alpha$-element abundances on their properties.  

{\em v)} We plan to apply a theoretical homogeneous approach in the calibration 
of TRGB distances, of secondary distance indicators and, eventually, in the 
estimate of the Hubble constant.

\begin{acknowledgements}

It is a pleasure to thank L. Rizzi for useful information  on distance 
determinations to external galaxies based on the TRGB method. We acknowledge 
the referee, Prof. M. Feast, for his pertinent comments and suggestions that
helped us to improve the content and the readability of the manuscript. 
We are very grateful to C. Jordi and V. Scowcroft for sending us their M31 
and NGC~598 Cepheid catalog in electronic form. We also acknowledge 
A.R. Walker for his suggestions and for a detailed reading of an early 
version of this manuscript.  
\end{acknowledgements}

%%%%%%%%%%%%%%%%%%%%%%%%%%%%%%%%%%%%%%%%%%%%%%%%%%%%%%%%%%%%%%%%%%%%%%%%%%%%%%%%%%%
%\begin{references}
%%%%%%%%%%%%%%%%%%%%%%%%%%%%%%%%%%%%%%%%%%%%%%%%%%%%%%%%%%%%%%%%%%%%%%%%%%%%%%%%%%
%\pagebreak

%%%%%%%%%%%%%%%%%%%%%%%%%%%%%%%%%%%%%%%%%%%%%%%%%%%%%%%%%%%%%%%%%%%%%%%%%%%%%%%
%				TABLE 1 
%%%%%%%%%%%%%%%%%%%%%%%%%%%%%%%%%%%%%%%%%%%%%%%%%%%%%%%%%%%%%%%%%%%%%%%%%%%%%%%
%\clearpage
\begin{deluxetable}{llcll}
\tablewidth{0pt}
\tabletypesize{\scriptsize}
\tablecaption{
Parameters of the computed pulsation models. From left to right: 
heavy element abundance ($Z$), helium content ($Y$), range in mass 
($\Delta M$, in solar units), luminosity ($L$), and mixing-length 
parameter ($l/H_p$). 
The adopted luminosity refers to the canonical Mass-Luminosity relation 
($L_{can}$) or to higher levels ($L_{over}$) predicted by evolutionary 
models accounting for mild convective core overshooting and/or affected 
by mass loss before or during the Cepheid phase (see text for details).
}\label{tbl-1}
\tablehead{
\colhead{$Z$}&
\colhead{$Y$}&
\colhead{$\Delta M$}&
\colhead{$L$}&
\colhead{ $l/H_p$} 
}
\startdata
0.004 &    0.25              &  3.5-11.0 & $L_{can}$, $L_{over}$& 1.5-1.8 \\
0.008 &    0.25              &  3.5-11.0 & $L_{can}$, $L_{over}$& 1.5-1.8 \\
0.01  &    0.26              &  5.0-11.0 & $L_{can}$            & 1.5-1.8 \\
0.02  &  0.25;0.26;0.28;0.31 &  5.0-11.0 & $L_{can}$, $L_{over}$& 1.5-1.8 \\
0.03  &   0.275;0.31;0.335   &  5.0-11.0 & $L_{can}$            &1.5   \\
0.04  &   0.25;0.29;0.33     &  5.0-11.0 & $L_{can}$            &1.5  \\
\enddata 
%\tablenotetext{a}{}   
\end{deluxetable}

%%%%%%%%%%%%%%%%%%%%%%%%%%%%%%%%%%%%%%%%%%%%%%%%%%%%%%%%%%%%%%%%%%%%%%%%%%%%%%%
%				TABLE 2 
%%%%%%%%%%%%%%%%%%%%%%%%%%%%%%%%%%%%%%%%%%%%%%%%%%%%%%%%%%%%%%%%%%%%%%%%%%%%%%%
%\clearpage
\begin{deluxetable}{lcccc}
\tablewidth{0pt}
\tabletypesize{\scriptsize}
\tablecaption{
Predicted slopes ($b$) of synthetic linear $P$-$L$ relations
$M_i$=$a$+$b$log$P$ for the chemical compositions listed in Table~1.  
The subscript $all$ refers to all the fundamental pulsators with
0.4$\le$log$P\le$2.0, while $short$ and $long$ refer to those with
log$P\le$1.0, and with log$P>$1.0, respectively. For each band, the 
dependence $\partial b_{all}/\partial \log (Z/X)$ is also listed.  
Note that for $Z\ge$0.02, the labeled ratio $Z/X$ is the mean value 
of the various helium abundances listed in Table~1.
}\label{tbl-2}
\tablehead{
\colhead{Band}&
\colhead{log$(Z/X)$}&
\colhead{$b_{all}$}&
\colhead{$b_{short}$}&
\colhead{$b_{long}$} 
}
\startdata
$M_B$   &   $-2.27$ &   $-$2.68$\pm$0.13 &   $-$3.04$\pm$0.16 & $-$2.00$\pm$0.15     \\
$M_B$   &   $-1.97$ &   $-$2.51$\pm$0.16    &   $-$2.89$\pm$0.12 & $-$1.92$\pm$0.15     \\
$M_B$   &   $-1.86$ &   $-$2.33$\pm$0.16 &   $-$2.86$\pm$0.12 & $-$1.62$\pm$0.15 \\
$M_B$   &   $-1.55$ &   $-$2.10$\pm$0.28 &   $-$2.23$\pm$0.21 & $-$1.66$\pm$0.13     \\
$M_B$   &   $-1.34$ &   $-$1.77$\pm$0.17    &   $-$1.81$\pm$0.19    &   $-$1.74$\pm$0.13    \\
$M_B$   &   $-1.22$ &   $-$1.77$\pm$0.17    &   $-$1.91$\pm$0.22    &   $-$1.82$\pm$0.15 \\

       \multicolumn{5}{c}{$\partial b_{all}(M_B)/\partial$log$(Z/X)$=$0.94\pm0.07$}    \\

$M_V$   & $-2.27$    &   $-$2.87$\pm$0.09    &   $-$3.22$\pm$0.08    &   $-$2.44$\pm$0.14     \\
$M_V$   & $-1.97$   &   $-$2.80$\pm$0.15    &   $-$2.97$\pm$0.10    &   $-$2.39$\pm$0.14     \\
$M_V$   & $-1.86$   &   $-$2.69$\pm$0.10 & $-$3.19$\pm$0.10 &   $-$2.14$\pm$0.14     \\
$M_V$   & $-1.55$   &   $-$2.51$\pm$0.24 & $-$2.60$\pm$0.18  &  $-$2.28$\pm$0.12     \\
$M_V$   & $-1.34$   &   $-$2.21$\pm$0.13    &   $-$2.33$\pm$0.15    &   $-$2.23$\pm$0.10     \\
$M_V$   & $-1.22$   &   $-$2.25$\pm$0.17    &   $-$2.51$\pm$0.22    &   $-$2.31$\pm$0.18 \\
       \multicolumn{5}{c}{$\partial b_{all}(M_V)/\partial$log$(Z/X)$=$0.67\pm0.09$}    \\
$M_I$   & $-2.27$    &   $-$3.00$\pm$0.07    &   $-$3.28$\pm$0.04    &   $-$2.76$\pm$0.10     \\
$M_I$   & $-1.97$   &   $-$2.90$\pm$0.13    &   $-$3.12$\pm$0.06    &   $-$2.70$\pm$0.10     \\
$M_I$   & $-1.86$   &   $-$2.91$\pm$0.06  & $-$3.30$\pm$0.09 &   $-$2.55$\pm$0.10     \\
$M_I$   & $-1.55$   &   $-$2.80$\pm$0.19  &  $-$2.88$\pm$0.13  & $-$2.61$\pm$0.10     \\
$M_I$   & $-1.34$   &   $-$2.56$\pm$0.09    &   $-$2.68$\pm$0.12    &   $-$2.61$\pm$0.05     \\
$M_I$   & $-1.22$  &   $-$2.61$\pm$0.13    &   $-$2.69$\pm$0.12    &   $-$2.63$\pm$0.15 \\
       \multicolumn{5}{c}{$\partial b_{all}(M_I)/\partial$log$(Z/X)$=$0.45\pm0.08$}    \\
$M_J$   & $-2.27$  &   $-$3.16$\pm$0.06    &   $-$3.33$\pm$0.05    &   $-$2.81$\pm$0.10     \\
$M_J$   & $-1.97$  &   $-$3.13$\pm$0.11    &   $-$3.30$\pm$0.04    &   $-$2.77$\pm$0.10     \\
$M_J$   & $-1.86$  &   $-$3.10$\pm$0.06    &   $-$3.37$\pm$0.10    &   $-$2.73$\pm$0.10    \\
$M_J$   & $-1.55$  &   $-$3.00$\pm$0.11    &   $-$3.06$\pm$0.16    &   $-$2.81$\pm$0.09     \\
$M_J$   & $-1.34$  &   $-$2.90$\pm$0.09    &   $-$2.94$\pm$0.10    &   $-$2.80$\pm$0.05     \\
$M_J$   & $-1.22$ &   $-$2.92$\pm$0.11    &   $-$2.95$\pm$0.08    &   $-$2.83$\pm$0.15   \\
       \multicolumn{5}{c}{$\partial b_{all}(M_J)/\partial$log$(Z/X)$=$0.27\pm0.03$}    \\
$M_K$   & $-2.27$  &   $-$3.19$\pm$0.09    &   $-$3.33$\pm$0.04    &   $-$3.04$\pm$0.09     \\
$M_K$   & $-1.97$   &   $-$3.28$\pm$0.09    &   $-$3.34$\pm$0.04    &   $-$3.02$\pm$0.09     \\
$M_K$   & $-1.86$    &   $-$3.31$\pm$0.03    &   $-$3.47$\pm$0.06    &   $-$3.05$\pm$0.09     \\
$M_K$   & $-1.55$    &   $-$3.22$\pm$0.15    &   $-$3.28$\pm$0.10    &   $-$3.08$\pm$0.08     \\
$M_K$   & $-1.34$   &   $-$3.16$\pm$0.06    &   $-$3.18$\pm$0.08    &   $-$3.13$\pm$0.05     \\
$M_K$   & $-1.22$  &   $-$3.16$\pm$0.09    &   $-$3.17$\pm$0.08    &   $-$3.12$\pm$0.13     \\
       \multicolumn{5}{c}{$\partial b_{all}(M_K)/\partial$log$(Z/X)$=$0.08\pm0.07$}    \\
\enddata 
\end{deluxetable}

%%%%%%%%%%%%%%%%%%%%%%%%%%%%%%%%%%%%%%%%%%%%%%%%%%%%%%%%%%%%%%%%%%%%%%%%%%%%%%%
%				TABLE 3 
%%%%%%%%%%%%%%%%%%%%%%%%%%%%%%%%%%%%%%%%%%%%%%%%%%%%%%%%%%%%%%%%%%%%%%%%%%%%%%%
\clearpage
\begin{deluxetable}{llccc}
\tablewidth{0pt}
\tabletypesize{\scriptsize}
\tablecaption{
Oxygen and iron abundances of the computed pulsation models, as 
derived for scaled-solar chemical compositions and by adopting 
the solar abundances from Asplund et al.\ (2004).
}\label{tbl-3}
\tablehead{
\colhead{$Z$}&
\colhead{$Y$}&
\colhead{log($Z/X$)}&
\colhead{12+log(O/H)}&
\colhead{[Fe/H]} 
}
\startdata
0.001 & 0.24 & $-$2.87	& 7.56&	$-$1.10\\
0.004 & 0.25 & $-$2.27	& 8.17&	$-$0.49\\
0.008 & 0.25 & $-$1.97	& 8.47&	$-$0.18\\
0.01  & 0.26 & $-$1.86	& 8.58&	$-$0.08\\
0.02  & 0.25 & $-$1.56	& 8.88&	+0.22\\
0.02  & 0.26 & $-$1.56	& 8.89&	+0.23\\
0.02  & 0.28 & $-$1.54	& 8.90&	+0.24\\
0.02  & 0.31 & $-$1.53	& 8.92&	+0.26\\
0.03  & 0.275  & $-$1.36	& 9.08&	+0.42\\
0.03  & 0.31  & $-$1.34	& 9.10&	+0.44\\
0.03  & 0.335  & $-$1.33	& 9.12&	+0.46\\
0.04  & 0.25 & $-$1.25	& 9.19&	 +0.53\\
0.04  & 0.29 & $-$1.22	& 9.22&	 +0.56\\
0.04  & 0.33 & $-$1.20	& 9.25&	 +0.59\\
\enddata 
%\tablenotetext{a}{}   
\end{deluxetable}

%%%%%%%%%%%%%%%%%%%%%%%%%%%%%%%%%%%%%%%%%%%%%%%%%%%%%%%%%%%%%%%%%%%%%%%%%%%%%%%
%				TABLE 4 
%%%%%%%%%%%%%%%%%%%%%%%%%%%%%%%%%%%%%%%%%%%%%%%%%%%%%%%%%%%%%%%%%%%%%%%%%%%%%%%
%\clearpage
\begin{deluxetable}{lcccccccc}
\tablewidth{0pt}
\tabletypesize{\scriptsize}
\tablecaption{
Published slopes ($b$) for observed $P$-$L$ relations. The
galaxy metallicity in the second column is the galaxy nebular 
abundance 12+log(O/H) in the new scale.
}\label{tbl-4}
\tablehead{
\colhead{Galaxy}&
\colhead{Met.}&
\colhead{log$P$}&
\colhead{$b(B)$}&
\colhead{$b(V)$}&
\colhead{$b(I)$}&
\colhead{$b(J)$}&
\colhead{$b(K_s)$}&
\colhead{Ref.\tablenotemark{a}} 
}
\startdata

 LMC &8.34& 0.4-2.0 &$-2.34\pm$0.04  & $-2.70\pm$0.03 &$-2.95\pm$0.02 &\ldots&\ldots& (1) \\
 LMC & "  & 0.4-1.0 & $-2.68\pm$0.08 & $-2.96\pm$0.06 &$-3.10\pm$0.04 &\ldots&\ldots& (") \\
 LMC & "  & 1.0-2.0 & $-2.15\pm$0.13 & $-2.57\pm$0.10 &$-2.82\pm$0.08 &\ldots&\ldots& (") \\
 LMC & "  & 0.4-2.0 & \ldots         & \ldots         &\ldots         &$-3.15\pm$0.05&$-3.28\pm$0.04& (2) \\
 LMC & "  & 0.4-1.7 &$-2.39\pm$0.04  & $-2.73\pm$0.03 &$-2.96\pm$0.02 &$-3.14\pm$0.03&$-3.23\pm$0.03& (3) \\
 LMC & "  & 0.4-1.0 & $-2.63\pm$0.07 & $-2.90\pm$0.05 &$-3.07\pm$0.04 &$-3.24\pm$0.04&$-3.29\pm$0.04& (4) \\
 LMC & "  & 1.0-1.7 & $-2.40\pm$0.19 & $-2.76\pm$0.14 &$-2.95\pm$0.10 &$-3.04\pm$0.15&$-3.21\pm$0.14& (") \\
 SMC &7.98& 0.4-1.7 & $-2.22\pm$0.05 & $-2.59\pm$0.05 &$-2.86\pm$0.04 &\ldots&\ldots& (5) \\
 SMC & "  & 0.4-1.0 & $-2.33\pm$0.10 & $-2.58\pm$0.09 &$-2.82\pm$0.07 &\ldots&\ldots& (") \\
 SMC & "  & 1.0-1.7 & $-2.35\pm$0.22 & $-2.79\pm$0.18 &$-3.06\pm$0.13 &\ldots&\ldots& (") \\
 MW  &8.60& 0.6-1.9 & $-2.69\pm$0.09 & $-3.09\pm$0.09 &$-3.35\pm$0.08 &\ldots&\ldots& (6)\\
 MW  & "  & 0.6-1.7 & $-2.29\pm$0.09 & $-2.68\pm$0.08 &$-2.98\pm$0.07 &$-3.19\pm$0.07&$-3.37\pm$0.06& (7) \\
 MW  & "  & 0.6-1.8 & \ldots         & $-2.60\pm$0.09 &\ldots         &\ldots&$-3.50\pm$0.08& (8) \\
 MW  & "  & 0.6-1.6 & \ldots         & $-2.43\pm$0.12 &$-2.81\pm$0.11 &\ldots&$-3.32\pm$0.12& (9) \\
 M31 &8.68& 0.4-1.6 & $-2.55$        & $-2.92\pm$0.21 &\ldots         &\ldots&\ldots& (10) \\

\enddata 
\tablenotetext{a}{References: 
(1) Sandage et al.\ (2004, STR04) using  OGLE (Udalski et al.\ 1999a) $BVI$ 
magnitudes plus additional data from different sources. 
(2) Persson et al.\ (2004, Pe04) using 2MASS near-infrared magnitudes. 
(3) Fouqu{\'e} et al.\ (2007, Fo07) using OGLE and Pe04 magnitudes plus additional near-infrared data. 
(4) Ngeow et al.\ (2008, NKN08) using the same dataset used by Fo07. 
(5) Sandage et al.\ (2009) using OGLE (Udalski et al.\ 1999b) $BVI$ magnitudes. 
(6) STR04 using either cluster or Baade-Wesselink distances to Galactic Cepheids. 
(7) Fo07 using revised Baade-Wesselink distances to selected Galactic Cepheids. 
(8) Groenewegen (2008, Gr08) using revised Baade-Wesselink distances to selected Galactic Cepheids. 
(9) Benedict et al. (2007, Be07) using trigonometric parallaxes to selected Galactic Cepheids. The 
K-band data are in the CIT near-infrared system.     
(10) Tammann et al.\ (2008, TSR08) using Cepheids with visual amplitude $A_V>$0.8 mag, observed by 
Vilardell et al.\ (2007), and individual reddenings based on the Galactic $P$-$C$ relation.
}   
\end{deluxetable}

%%%%%%%%%%%%%%%%%%%%%%%%%%%%%%%%%%%%%%%%%%%%%%%%%%%%%%%%%%%%%%%%%%%%%%%%%%%%%%%
%				TABLE 5 
%%%%%%%%%%%%%%%%%%%%%%%%%%%%%%%%%%%%%%%%%%%%%%%%%%%%%%%%%%%%%%%%%%%%%%%%%%%%%%%
%\clearpage
\begin{deluxetable}{lcccccccc}
\tablewidth{0pt}
\tabletypesize{\scriptsize}
\tablecaption{
Slopes ($b$) of the observed $P$-$L$ relations determined
in the present paper. The metallicity listed in the second 
column is the galaxy nebular abundance 12+log(O/H) in the 
new metallicity scale. 
}\label{tbl-5}
\tablehead{
\colhead{Galaxy}&
\colhead{Met.}&
\colhead{log$P$}&
\colhead{$b(B)$}&
\colhead{$b(V)$}&
\colhead{$b(I)$}&
\colhead{$b(J)$}&
\colhead{$b(K_s)$}&
\colhead{Notes\tablenotemark{a}} 
}
\startdata

 SMC &7.98& 0.5-2.0 & $-2.45\pm$0.05 & $-2.80\pm$0.08 &$-3.02\pm$0.07 &\ldots&\ldots& (1) \\
 SMC & "  & 0.5-1.0 & $-2.28\pm$0.15 & $-2.56\pm$0.12 &$-2.80\pm$0.09 &\ldots&\ldots& (") \\
 SMC &  " & 1.0-2.0 & $-2.50\pm$0.17 & $-2.89\pm$0.13 &$-3.15\pm$0.12 &$-3.41\pm$0.19&$-3.56\pm$0.18& (") \\
 MW  &8.60& 0.6-1.8 & $-2.29\pm$0.10 & $-2.71\pm$0.10 &$-3.04\pm$0.09 &$-3.29\pm$0.09&$-3.45\pm$0.09& (2) \\
 MW  &"   & 0.6-1.8 & $-2.32\pm$0.10 & $-2.74\pm$0.12 &$-3.07\pm$0.11 &$-3.31\pm$0.09&$-3.47\pm$0.09& (3) \\
 M31 &8.68& 0.5-1.6 & $-2.29\pm$0.15 & $-2.72\pm$0.11 &\ldots  &\ldots &\ldots & (3) \\

\enddata 
\tablenotetext{a}{Notes: 
(1) OGLE (Udalski et al.\ 1999b) magnitudes plus additional Cepheids
from Caldwell \& Coulson (1984, $BVI$) and Laney \& Stobie (1986, $JK$). 
(2) Entire sample of Galactic Cepheids with distances measured by Fo07. 
(3) Entire sample of Galactic Cepheids with distances measured by Gr08. 
(4) Cepheids with visual amplitude $A_V>$0.8 mag, observed by 
Vilardell et al.\ (2007), but without reddening correction.
}   
\end{deluxetable}

%%%%%%%%%%%%%%%%%%%%%%%%%%%%%%%%%%%%%%%%%%%%%%%%%%%%%%%%%%%%%%%%%%%%%%%%%%%%%%%
%				TABLE 6 
%%%%%%%%%%%%%%%%%%%%%%%%%%%%%%%%%%%%%%%%%%%%%%%%%%%%%%%%%%%%%%%%%%%%%%%%%%%%%%%
%\clearpage
\begin{deluxetable}{lcccc}
\tablewidth{0pt}
\tabletypesize{\scriptsize}
\tablecaption{
Slopes $b(V)$ and $b(I)$ for observed $P$-$L$ relations 
provided by Tammann et al.\ (2008, TSR08) and by Saha et al.\ (2006, STT06).
The galaxy metallicity in the second column is the nebular oxygen abundance
12+log(O/H) in the new scale.
}\label{tbl-6}
\tablehead{
\colhead{Galaxy}&
\colhead{Met.}&
\colhead{$b(V)$}&
\colhead{$b(I)$}&
\colhead{Notes\tablenotemark{a}} 
}
\startdata

\multicolumn{5}{c}{TSR08} \\
N3351 &   8.85    &   $-3.12\pm$0.39 & $-$3.38 &    \\
N4321 &   8.74    &   $-3.17\pm$0.34 & $-$3.43  &  \\
N0224     &   8.68    &   $-2.92\pm$0.21 & -       & \\
MW      &   8.60    &   $-3.09\pm$0.09 & $-$3.35  &  \\
LMC     &   8.34    &   $-2.70\pm$0.03 & $-$2.95   &  \\
N6822 &   8.14    &   $-2.49\pm$0.01 & $-$2.81  &   \\
N3109 &   8.06    &   $-2.13\pm$0.18 & $-$2.40  &    \\
SMC     &   7.98    &   $-2.59\pm$0.05 & $-$2.86  &  \\
IC1613  &   7.86    &   $-2.67\pm$0.12 & $-$2.80  &    \\
WLM     &   7.74    &   $-2.52\pm$0.15 & $-$2.74  &   \\
Sext(A+B)  &   7.52    &   $-1.59\pm$0.39 & $-$1.47  & (1)  \\
M-P     &   8.36    &   $-2.69\pm$0.12 & $-$2.97   & (2)  \\
M-R     &   8.75    &   $-2.88\pm$0.13 & $-$3.13  &  (3) \\
\multicolumn{5}{c}{STT06} \\
N3351  &   8.85    &   $-$2.52 &       $-$3.01  &   \\
N4548  &   8.85    &   $-$1.35 &       $-$2.31  &  \\
N3627  &   8.80    &   $-$2.28 &       $-$2.64  &   \\
N4535  &   8.77    &   $-$2.85 &       $-$3.05  &  \\
N4321  &   8.74    &   $-$3.02 &       $-$3.32  &  \\
N5457i &   8.70    &   $-$2.36 &       $-$2.68  & (4) \\
N1425   &   8.67    &   $-$1.85 &       $-$2.13  &  \\
N3198  &   8.43    &   $-$2.36 &       $-$2.60  &  \\
N0925   &   8.40    &   $-$2.78 &       $-$2.88  &  \\
N2541  &   8.37    &   $-$2.53 &       $-$3.26  &  \\
N1326A &   8.37    &   $-$2.75 &       $-$2.99  &  \\
N0300   &   8.35    &   $-$2.79 &       $-$2.95  &  \\
N3319  &   8.28    &   $-$2.41 &       $-$3.11  &  \\
N5457o &   8.23    &   $-$2.13 &       $-$2.71  & (5) \\

\enddata 
\tablenotetext{a}{Notes: 
(1) The entire sample of Cepheids in the two galaxies Sextans A and Sextans B 
are combined in a single relation. 
(2) Mean values of the metal-poor [12+log(O/H)$<$8.45] galaxies from STT06. 
(3) Mean values of the metal-rich [12+log(O/H)$>$8.65] galaxies from STT06. 
(4) Inner field Cepheids.
(5) Outer field Cepheids.
}   
\end{deluxetable}

%%%%%%%%%%%%%%%%%%%%%%%%%%%%%%%%%%%%%%%%%%%%%%%%%%%%%%%%%%%%%%%%%%%%%%%%%%%%%%%
%				TABLE 7 
%%%%%%%%%%%%%%%%%%%%%%%%%%%%%%%%%%%%%%%%%%%%%%%%%%%%%%%%%%%%%%%%%%%%%%%%%%%%%%%
%\clearpage
\begin{deluxetable}{llll}
\tablewidth{0pt}
\tabletypesize{\scriptsize}
\tablecaption{Galaxy metal abundances in the notation 12+log(O/H). The
values marked with an asterisk have been estimated using the polynomial 
regression presented in Fig.~7 (see text for more details).}\label{tbl-7}
\tablehead{
\colhead{Galaxy}&
\colhead{Ref.\tablenotemark{a}}&
\colhead{Met$_{old}$}&
\colhead{Met$_{new}$} 
}
\startdata
IC1613	&	Sa04	&	7.86$\pm$0.50	&	7.86	\\		
IC4182	&	STT06	&	8.40$\pm$0.20	&	8.20	\\		
LMC	&	Sa04	&	8.50$\pm$0.08	&	8.34	\\		
N0224	&	STT06	&	8.98$\pm$0.15	&	8.68	\\
N0300	&	STT06	&	8.35$\pm$0.15	&	8.35	\\	
\enddata 
\tablenotetext{~}{NOTE. -- Table~7 is presented in its entirety
in the electronic edition of the manuscript. A portion is shown
here for guidance regarding its form and content.}
\tablenotetext{a}{References: Scowcroft et al.\ (2009); 
Riess et al.\ (2009, Ri09);  McCommas et al.\ (2009, hereinafter McC09).}   
\end{deluxetable}

%%%%%%%%%%%%%%%%%%%%%%%%%%%%%%%%%%%%%%%%%%%%%%%%%%%%%%%%%%%%%%%%%%%%%%%%%%%%%%%
%                               TABLE 8 
%%%%%%%%%%%%%%%%%%%%%%%%%%%%%%%%%%%%%%%%%%%%%%%%%%%%%%%%%%%%%%%%%%%%%%%%%%%%%%%
%\clearpage
\begin{deluxetable}{lccccl}
\tablewidth{0pt}
\tabletypesize{\scriptsize}
\tablecaption{
Slopes $b(V)$ and $b(I)$ of observed $P$-$L$ relations estimated using
Cepheids with 0.5$\le$log$P\le$2.0. The subscripts $i$ and $o$ refer either
to the inner field or to the outer field Cepheids. The galaxies with a
Cepheid number smaller than 20 are marked with an asterisk. Column 5 gives
the slope $\beta(WVI)$ of the observed $P$-$WVI$ relations discussed in
Sect.~4.
}\label{tbl-8}
\tablehead{
\colhead{Galaxy}&
\colhead{Met.}&
\colhead{$b(V)$}&
\colhead{$b(I)$}&
\colhead{$\beta(WVI)$}&
\colhead{Ref.\tablenotemark{a}}
}
\startdata
IC1613  &   7.86  & $-$2.72$\pm$0.13 &$-$2.92$\pm$0.10 & $-3.34\pm$0.06 &       (1,2)   \\
IC4182  &       8.20    & $-$2.80$\pm$0.17      &       $-$3.23$\pm$0.18& $-3.93\pm$0.26 &      Ka03    \\
IC4182  &       8.20    & $-$2.60$\pm$0.17      &       $-$2.93$\pm$0.18& $-3.43\pm$0.26 &      KPr     \\
IC4182  &       8.20    & $-$2.73$\pm$0.16      &       $-$3.15$\pm$0.17& $-3.81\pm$0.19 &      STT06   \\
LMC       &     8.34    & $-$2.70$\pm$0.03      &       $-$2.95$\pm$0.02& $-3.37\pm$0.03 &      STR04   \\
MW        &     8.60    & $-$2.71$\pm$0.10      &       $-$3.04$\pm$0.09& $-3.52\pm$0.12 &      Table~5 \\
N0055     &\ldots & $-$2.29$\pm$0.15    &       $-$2.56$\pm$0.12        & $-2.90\pm$0.13 &      (3,4)   \\
N0224   &8.68   & $-$2.72$\pm$0.11      &\ldots & \ldots &      Table~5         \\
N0247*& \ldots    & $-$2.34$\pm$0.19    &       $-$2.71$\pm$0.16        & $-3.39\pm$0.35 &      (5,6,7)\\
N0300   &       8.35    & $-$3.00$\pm$0.11      &       $-$3.12$\pm$0.12        & $-3.29\pm$0.12 &      (8,9)\\
\enddata
\tablenotetext{~}{NOTE. -- Table~8 is presented in its entirety
in the electronic edition of the manuscript. A portion is shown
here for guidance regarding its form and content.}
\tablenotetext{a}{References:
[KP], galaxies download from the KP web page.
[STT06], SNP galaxies revised by Saha et al.\ (2006).
[Ka03], KP and SNP galaxies revised by Kanbur et al.\ (2003).
[KPr], KP and SNP galaxies revised by the KP group.
[(1)], Udalski et al.\ (2001).
[(2)], Pietrzynski et al.\ (2006a).
[(3)], Pietrzynski et al.\ (2006b).
[(4)], Gieren et al.\ (2008a).
[(5)], Garcia-Varela et al.\ (2008).
[(6)], Madore et al.\ (2009).
[(7)], Gieren et al.\ (2009).
[(8)], Gieren et al.\ (2004).
[(9)], Gieren et al.\ (2005).
[(10)], Scowcroft et al.\ (2009).
[(11)], Riess et al.\ (2009).
[(12)], Leonard et al.\ (2003).
[(13)], Macri et al.\ (2001).
[(14)], McCommas et al.\ (2009).
[(15)], Pietrzynski et al.\ (2006c).
[(16)], Soszynski et a.\ (2006).
[(17)], Tanvir, Ferguson \& Shanks (1999).
[(18)], Riess et al.\ (2005).
[(19)], Stetson \& Gibson (2001).
[(20)], Macri et al.\ (2006).
[(21)], Newman et al.\ (2001).
[(22)], Gibson \& Stetson (2001).
[(23)], Ferrarese et al.\ (2007).
[(24)], Thim et al.\ (2003).
[(25)], Pietrzynski et al.\ (2004).
[(26)], Gieren et al.\ (2006).
[(27)], Piotto, Capaccioli \& Pellegrini (1994).
[(28)], Dolphin et al.\ (2003).
[(29)], Pietrzynski et al.\ (2007).
[(30)], Gieren et al.\ (2008b).
}
\end{deluxetable}

%%%%%%%%%%%%%%%%%%%%%%%%%%%%%%%%%%%%%%%%%%%%%%%%%%%%%%%%%%%%%%%%%%%%%%%%%%%%%%%
%				TABLE 9 
%%%%%%%%%%%%%%%%%%%%%%%%%%%%%%%%%%%%%%%%%%%%%%%%%%%%%%%%%%%%%%%%%%%%%%%%%%%%%%%
%\clearpage

\begin{deluxetable}{lccc}
\tablewidth{0pt}
\tabletypesize{\scriptsize}
\tablecaption{
Predicted slopes $\beta(W)$ of the $P$-$W$ relations given by the 
intensity-weighted mean magnitudes of fundamental pulsators with
0.5$\le$log$P\le$2.0 and $Z$=0.001-0.04. The metallicity in the first 
column is the oxygen abundance 12+log(O/H) listed in Table~3.
}\label{tbl-9}
\tablehead{
\colhead{Met.}&
\colhead{$\beta(WBV)$}&
\colhead{$\beta(WVI)$}&
\colhead{$\beta(WJK_s)$} 
}
\startdata
7.56	&$-$3.80$\pm$0.03 &$-$3.35$\pm$0.03 &	$-$3.37$\pm$0.03  \\
8.17	&$-$3.83$\pm$0.02 &$-$3.37$\pm$0.03 &	$-$3.39$\pm$0.03  \\
8.47	&$-$3.81$\pm$0.03 &$-$3.31$\pm$0.04 &	$-$3.38$\pm$0.03  \\
8.58	&$-$3.80$\pm$0.03 &$-$3.24$\pm$0.04 &	$-$3.37$\pm$0.04  \\
8.89 	&$-$3.82$\pm$0.03 &$-$3.22$\pm$0.03 &	$-$3.36$\pm$0.03  \\
9.10 	&$-$3.77$\pm$0.03 &$-$3.20$\pm$0.03 &	$-$3.31$\pm$0.03  \\
9.22	&$-$3.76$\pm$0.03 &$-$3.23$\pm$0.03 &	$-$3.32$\pm$0.03  \\
\enddata 
%\tablenotetext{a}{}   
\end{deluxetable}

%%%%%%%%%%%%%%%%%%%%%%%%%%%%%%%%%%%%%%%%%%%%%%%%%%%%%%%%%%%%%%%%%%%%%%%%%%%%%%%
%				TABLE 10 
%%%%%%%%%%%%%%%%%%%%%%%%%%%%%%%%%%%%%%%%%%%%%%%%%%%%%%%%%%%%%%%%%%%%%%%%%%%%%%%
%\clearpage
\begin{deluxetable}{lccc}
\tablewidth{0pt}
\tabletypesize{\scriptsize}
\tablecaption{
Observed slopes of the $P$-$WBV$ and $P$-$WJK_s$ relations, 
as derived in the present paper for Cepheids with 0.5$\le$log$P\le$2.0.
}\label{tbl-10}
\tablehead{
\colhead{Galaxy}&
\colhead{Met.}&
\colhead{$\beta(WBV)$}&
\colhead{$\beta(WJK_s)$} 
}
\startdata
IC1613 	               & 7.86  &\ldots           & $-$3.27$\pm$0.19  \\
IC1613\tablenotemark{a}& 7.86  &$-$3.70$\pm$0.15 &\ldots\\
LMC    	               & 8.34  &$-$3.82$\pm$0.04 & $-$3.45$\pm$0.06  \\
MW     	               & 8.60  &$-$4.04$\pm$0.12 & $-$3.54$\pm$0.10  \\
N0055  	               &\ldots &\ldots	         & $-$3.19$\pm$0.18  \\
N0224	               & 8.68  &$-$3.99$\pm$0.10 &\ldots  \\
N0247  	               &\ldots &\ldots           &$-$3.26$\pm$0.15  \\
N0300  	               & 8.35  &$-$4.04$\pm$0.28 &$-$3.12$\pm$0.25  \\
N0598i                 & 8.58  &$-$3.23$\pm$0.30 &\ldots\\
N0598o                 & 8.21  &$-$3.08$\pm$0.30 &\ldots\\
N4258i 	               & 8.64  &$-$3.77$\pm$0.10 &\ldots\\
N4258o 	               & 8.50  &$-$3.97$\pm$0.10 &\ldots\\
N6822 	               & 8.14  &\ldots           &$-$3.12$\pm$0.14\\
SextA 	               & 7.49  &$-$2.41$\pm$0.38 &\ldots\\
SextB                  & 7.56  &$-$3.96$\pm$0.40 &\ldots\\   
SMC    	               & 7.98  &$-$3.94$\pm$0.05 &$-$3.29$\pm$0.16  \\
WLM	 	       & 7.74  &\ldots		 &$-$3.17$\pm$0.19\\
\enddata 
\tablenotetext{a}{Data from Antonello et al.\ (2006)}   
\end{deluxetable}

%%%%%%%%%%%%%%%%%%%%%%%%%%%%%%%%%%%%%%%%%%%%%%%%%%%%%%%%%%%%%%%%%%%%%%%%%%%%%%%
%				TABLE 11 
%%%%%%%%%%%%%%%%%%%%%%%%%%%%%%%%%%%%%%%%%%%%%%%%%%%%%%%%%%%%%%%%%%%%%%%%%%%%%%%
%\clearpage
\begin{deluxetable}{llccc}
\tablewidth{0pt}
\tabletypesize{\scriptsize}
\tablecaption{
Mean LMC-relative true distance moduli of canonical fundamental pulsators with
0.5$\le$log$P\le$2.0 and $Z$=0.001-0.04 based on the intensity-averaged magnitudes 
of the pulsators and on the observed LMC $P$-$W$ relations discussed in the text. 
The metallicity in the first column is the oxygen abundance 12+log(O/H) listed in 
Table~3.
}\label{tbl-11}
\tablehead{
\colhead{Met.}&
\colhead{$Y$}&
\colhead{$\delta\mu_0(WBV)$}&
\colhead{$\delta\mu_0(WVI)$}&
\colhead{$\delta\mu_0(WJK_s)$} 
}
\startdata
\multicolumn{5}{c}{log$L/L_{can}$=0} \\
7.56  & 0.24 &$-$18.16$\pm$0.05 &$-$18.68$\pm$0.11 &	$-$18.65$\pm$0.10  \\
8.17	& 0.25 &$-$18.46$\pm$0.06 &$-$18.75$\pm$0.11 &	$-$18.75$\pm$0.09  \\
8.48	& 0.25 &$-$18.72$\pm$0.05 &$-$18.77$\pm$0.09 &	$-$18.79$\pm$0.08  \\
8.58	& 0.26 &$-$18.72$\pm$0.07 &$-$18.73$\pm$0.12 &	$-$18.74$\pm$0.10  \\
8.88  & 0.25 &$-$18.99$\pm$0.08 &$-$18.70$\pm$0.10 &	$-$18.73$\pm$0.08  \\
8.89	& 0.26 &$-$18.97$\pm$0.07 &$-$18.74$\pm$0.10 &	$-$18.75$\pm$0.08  \\
8.90	& 0.28 &$-$18.94$\pm$0.08 &$-$18.67$\pm$0.12 &  $-$18.73$\pm$0.08\\
8.92	& 0.31 &$-$18.90$\pm$0.04 &$-$18.65$\pm$0.11 &  $-$18.64$\pm$0.08\\
9.08  & 0.275&$-$19.05$\pm$0.07 &$-$18.72$\pm$0.09 &	$-$18.66$\pm$0.09  \\
9.10	& 0.31 &$-$19.00$\pm$0.07 &$-$18.64$\pm$0.09 &	$-$18.59$\pm$0.08  \\
9.12	& 0.335&$-$18.98$\pm$0.07 &$-$18.64$\pm$0.08 &	$-$18.57$\pm$0.08  \\
9.19	& 0.25 &$-$19.20$\pm$0.08 &$-$18.82$\pm$0.07 &	$-$18.70$\pm$0.05  \\
9.22	& 0.29 &$-$19.13$\pm$0.06 &$-$18.77$\pm$0.06 &	$-$18.65$\pm$0.07  \\
9.25	& 0.33 &$-$19.07$\pm$0.08 &$-$18.71$\pm$0.06 &	$-$18.58$\pm$0.08  \\
\multicolumn{5}{c}{log$L/L_{can}$=0.20} \\
8.17	& 0.25 &$-$18.27$\pm$0.06 &$-$18.55$\pm$0.11 &	$-$18.57$\pm$0.09  \\
8.48	& 0.25 &$-$18.54$\pm$0.04 &$-$18.57$\pm$0.09 &	$-$18.61$\pm$0.08  \\
8.88  & 0.28 &$-$18.75$\pm$0.07 &$-$18.47$\pm$0.10 &	$-$18.55$\pm$0.08  \\
\enddata 
%\tablenotetext{a}{}   
\end{deluxetable}

%%%%%%%%%%%%%%%%%%%%%%%%%%%%%%%%%%%%%%%%%%%%%%%%%%%%%%%%%%%%%%%%%%%%%%%%%%%%%%%
%				TABLE 12 
%%%%%%%%%%%%%%%%%%%%%%%%%%%%%%%%%%%%%%%%%%%%%%%%%%%%%%%%%%%%%%%%%%%%%%%%%%%%%%%
%\clearpage
\begin{deluxetable}{llccc}
\tablewidth{0pt}
\tabletypesize{\scriptsize}
\tablecaption{
Internal differences among the LMC-relative true distance moduli listed in Table~11.
}\label{tbl-12}
\tablehead{
\colhead{Met.}&
\colhead{$Y$}&
\colhead{$\Delta(WBV-WVI)$}&
\colhead{$\Delta(WBV-WJK_s)$}&
\colhead{$\Delta(WVI-WJK_s)$} 
}
\startdata
\multicolumn{5}{c}{log$L/L_{can}$=0} \\
7.56  & 0.24 &+0.42$\pm$0.10 &+0.49$\pm$0.10 &	+0.03$\pm$0.10  \\
8.17	& 0.25 &+0.26$\pm$0.11 &+0.29$\pm$0.08 &	0$\pm$0.11  \\
8.48	& 0.25 &+0.04$\pm$0.09 &+0.07$\pm$0.07 &	+0.03$\pm$0.08  \\
8.58	& 0.26 &+0.01$\pm$0.11 &+0.02$\pm$0.09 &	+0.01$\pm$0.11  \\
8.88  & 0.25 &$-$0.29$\pm$0.09 &$-$0.26$\pm$0.07 &	+0.03$\pm$0.08  \\
8.89	& 0.26 &$-$0.23$\pm$0.10 &$-$0.22$\pm$0.07 &	+0.01$\pm$0.10  \\
8.90	& 0.28 &$-$0.29$\pm$0.11 &$-$0.21$\pm$0.07 &  +0.06$\pm$0.10\\
8.92	& 0.31 &$-$0.25$\pm$0.10 &$-$0.26$\pm$0.07 &  $-$0.01$\pm$0.10\\
9.08  & 0.275&$-$0.34$\pm$0.08 &$-$0.39$\pm$0.06 &	$-$0.06$\pm$0.08  \\
9.10	& 0.31 &$-$0.36$\pm$0.08 &$-$0.41$\pm$0.07 &	$-$0.05$\pm$0.09  \\
9.12	& 0.335&$-$0.34$\pm$0.08 &$-$0.41$\pm$0.07 &	$-$0.07$\pm$0.09  \\
9.19	& 0.25 &$-$0.37$\pm$0.06 &$-$0.50$\pm$0.05 &	$-$0.12$\pm$0.06  \\
9.22	& 0.29 &$-$0.36$\pm$0.06 &$-$0.50$\pm$0.06 &	$-$0.14$\pm$0.07  \\
9.25	& 0.33 &$-$0.36$\pm$0.06 &$-$0.49$\pm$0.07 &	$-$0.16$\pm$0.07  \\
\multicolumn{5}{c}{log$L/L_{can}$=0.20} \\
8.17	& 0.25 &+0.26$\pm$0.10 &+0.30$\pm$0.09 &	+0.02$\pm$0.09  \\
8.48	& 0.25 &+0.03$\pm$0.08 &+0.08$\pm$0.09 &	+0.05$\pm$0.09  \\
8.88  & 0.28 &$-$0.28$\pm$0.08 &$-$0.20$\pm$0.07 &	+0.08$\pm$0.09  \\
\enddata 
%\tablenotetext{a}{}   
\end{deluxetable}

%%%%%%%%%%%%%%%%%%%%%%%%%%%%%%%%%%%%%%%%%%%%%%%%%%%%%%%%%%%%%%%%%%%%%%%%%%%%%%%
%				TABLE 13 
%%%%%%%%%%%%%%%%%%%%%%%%%%%%%%%%%%%%%%%%%%%%%%%%%%%%%%%%%%%%%%%%%%%%%%%%%%%%%%%
%\clearpage
\begin{deluxetable}{lcccc}
\tablewidth{0pt}
\tabletypesize{\scriptsize}
\tablecaption{
Mean LMC-relative true distance moduli of external galaxies based on 
Cepheids with 0.5$\le$log$P\le$2.0 and  the LMC $P$-$W$ relations discussed 
in the text. The galaxies are ordered by metal abundance.
}\label{tbl-13}
\tablehead{
\colhead{Galaxy}&
\colhead{Met.}&
\colhead{$\delta\mu_0(WBV)$}&
\colhead{$\delta\mu_0(WVI)$}&
\colhead{$\delta\mu_0(WJK_s)$} 
}
\startdata
SextA        & 7.49  & 7.57$\pm$0.19  & 7.18$\pm$0.24   & \ldots\\
SextB        & 7.56  & 7.69$\pm$0.51  & \ldots   & \ldots\\
WLM	     & 7.74  &\ldots          &6.58$\pm$0.11    &6.40$\pm$0.18\\
IC1613\tablenotemark{a} 	 & 7.86  &6.17$\pm$0.27    & 5.93$\pm$0.13   & \ldots\\
IC1613\tablenotemark{b} 	 & 7.86  &\ldots       & 5.72$\pm$0.13   & 5.78$\pm$0.22\\
SMC          & 7.98  &0.73$\pm$0.23    & 0.45$\pm$0.12   & 0.41$\pm$0.20\\
N6822 	 & 8.14  &\ldots  & 4.88$\pm$0.22   & 4.84$\pm$0.18\\
N0598o  	 & 8.21  &6.17$\pm$0.31    & 6.15$\pm$0.19   & \ldots\\
N0300  	 & 8.35  &8.14$\pm$0.32   & 7.98$\pm$0.29   & 7.87$\pm$0.28  \\
N4258o 	 & 8.50  &10.99$\pm$0.21   & 10.90$\pm$0.13  & \ldots\\
N0598i  	 & 8.58  &5.89$\pm$0.30    & 5.91$\pm$0.27   &\ldots  \\
N4258i 	 & 8.64  &10.65$\pm$0.21   &  10.68$\pm$0.21 &\ldots\\
N0224   	 & 8.68  & 5.66$\pm$0.25       & \ldots& \ldots\\
N0055   	 & \ldots &\ldots & 8.02$\pm$0.38   & 7.91$\pm$0.28  \\
N0247  	 & \ldots  &\ldots              & 9.35$\pm$0.35   & 9.18$\pm$0.20  \\
\enddata 
\tablenotetext{a}{Antonello et al.\ (2006).}   
\tablenotetext{b}{Udalski et al.\ (2001), Pietrzynski et al.\ (2006a).}   
\end{deluxetable}

%%%%%%%%%%%%%%%%%%%%%%%%%%%%%%%%%%%%%%%%%%%%%%%%%%%%%%%%%%%%%%%%%%%%%%%%%%%%%%%
%				TABLE 14 
%%%%%%%%%%%%%%%%%%%%%%%%%%%%%%%%%%%%%%%%%%%%%%%%%%%%%%%%%%%%%%%%%%%%%%%%%%%%%%%
%\clearpage
\begin{deluxetable}{lc}
\tablewidth{0pt}
\tabletypesize{\scriptsize}
\tablecaption{
TRGB distance moduli provided by Rizzi et al.\ (2007, Ri07). 
The galaxies are ordered by metal abundance. 
}\label{tbl-14}
\tablehead{
\colhead{Galaxy}&
\colhead{Ri07} 
}
\startdata
SextA	&25.79$\pm$0.06	\\
SextB	&25.79$\pm$0.04	\\			
WLM	&24.93$\pm$0.04	\\	
IC1613  &24.37$\pm$0.05 	\\	
SMC	&18.98	        \\
N3109	&25.56$\pm$0.05	\\
IC4182  & 28.23$\pm$0.05	\\		
N0598	& 24.71$\pm$0.04	\\
N5457	& 29.34$\pm$0.09	\\
LMC	& 18.57	        \\
N0300	& 26.48$\pm$0.04	\\		
N3031	& 27.69$\pm$0.04	\\
N3621	& 29.26$\pm$0.12	\\
N4258	& 29.42$\pm$0.06	\\		
N0224	& 24.37$\pm$0.10	\\	
N3351	& 29.92$\pm$0.05	\\
N5128	& 27.72$\pm$0.04	\\				
\enddata 
%\tablenotetext{a}{}   
\end{deluxetable}

%%%%%%%%%%%%%%%%%%%%%%%%%%%%%%%%%%%%%%%%%%%%%%%%%%%%%%%%%%%%%%%%%%%%%%%%%%%%%%%
%                               TABLE 15 
%%%%%%%%%%%%%%%%%%%%%%%%%%%%%%%%%%%%%%%%%%%%%%%%%%%%%%%%%%%%%%%%%%%%%%%%%%%%%%%
%\clearpage
\begin{deluxetable}{llccccc}
\tablewidth{0pt}
\tabletypesize{\scriptsize}
\tablecaption{
Distance moduli $\mu_0(WVI)$ for
the galaxies listed in Table~8. The number in square brackets in column (2)
is the LMC distance modulus adopted in the original paper.
}\label{tbl-15}
\tablehead{
\colhead{Galaxy}&
\colhead{Ref.[$\mu_{0,LMC}]$}&
\colhead{Source}&
\colhead{Fr01}&
\colhead{Ka03}&
\colhead{STT06}&
\colhead{Our}
}
\startdata
IC1613\tablenotemark{a}&  &\ldots                       &24.19$\pm0.15$   &\ldots               &\ldots  & \ldots                       \\
IC1613  &(1)[18.50]       &24.17$\pm0.07$       &\ldots                 &\ldots         &\ldots         & 24.22$\pm0.10$                \\
IC4182  &       Ka03      &     \ldots          &       \ldots                  & 28.33$\pm0.13$        &\ldots & 28.35$\pm0.23$        \\
IC4182  &       KPr       &     \ldots          & 28.28$\pm0.06$  &     \ldots  &\ldots       & 28.25$\pm0.18$  \\
IC4182  &       STT06     &     \ldots          &       \ldots                  &       \ldots  & 28.28$\pm0.10$& 28.29$\pm0.18$        \\
LMC     &       STR04     &18.50        & 18.50 &18.50  & 18.50 &18.50  \\
N0055   &   (3)[18.50]    &26.40$\pm0.05$       &       \ldots                  &       \ldots  &\ldots                 & 26.47$\pm0.18$\\
N0224   &       MF05      &     \ldots          & 24.38$\pm0.05$  &     \ldots  & 24.35&\ldots                          \\
N0247   &(5,6)[18.50]     &27.80$\pm0.09$       &       \ldots                  &       \ldots  &\ldots         & 27.84$\pm0.18$  \\
N0300   &(8,9)[18.50]     &26.43$\pm0.06$       & 26.53$\pm0.07$  &     \ldots  & 26.44& 26.44$\pm0.16$         \\
\enddata
\tablenotetext{~}{NOTE. -- Table~15 is presented in its entirety
in the electronic edition of the manuscript. A portion is shown
here for guidance regarding its form and content.}
\tablenotetext{a}{Data from Freedman (1988).}
\tablenotetext{b}{Data from Freedman et al.\ (1991).}
\tablenotetext{c}{Distance modulus from Sandage \& Tammann (2008).}
\end{deluxetable}

\end{document}